\documentclass[aps,prd,showpacs]{revtex4}
\usepackage{epsf,epsfig,graphicx,axodraw}
\begin{document}

\begin{flushright}
SI-HEP-2012-02 \\
\end{flushright}

\date{\today}

\title{Next-to-leading-order corrections to $B \to \pi$ form factors in $k_T$ factorization}

\author{Hsiang-nan Li$^{a,b,c}$, Yue-Long Shen$^{d}$,  Yu-Ming Wang$^{e}$ }
\affiliation{{\it \small $^a$Institute of Physics, Academia Sinica,
Taipei, Taiwan 115, Republic of China }\\ {\it \small $^b$Department
of Physics, National Cheng-Kung University, Tainan, Taiwan 701,
Republic of China}\\ {\it \small $^c$Department of Physics, National
Tsing-Hua University, Hsinchu, Taiwan 300, Republic of China}\\
{\it \small $^d$College of Information Science and Engineering,
Ocean University of China, Qingdao, Shandong 266100, P.R. China} \\
{\it \small $^e$ Theoretische Elementarteilchenphysik,  Naturwissenschaftlich Techn. Fakult\"at \\
Universi\"at Siegen, 57068 Siegen, Germany}}

\begin{abstract}

We calculate next-to-leading-order (NLO) corrections to the
$B\to\pi$ transition form factors at leading twist in the $k_T$
factorization theorem. Light partons off-shell by $k_T^2$ are
considered in the quark diagrams, in the effective diagrams for
the $B$ meson wave function defined with the effective heavy-quark
field, and in the effective diagrams for the pion wave function.
It is explicitly demonstrated that the infrared logarithms $\ln
k_T^2$ cancel between the above sets of diagrams, as deriving the
$k_T$-dependent NLO hard kernel from their difference. The
infrared finiteness of the hard kernel confirms the application
of the $k_T$ factorization theorem to $B$ meson semileptonic
decays. The NLO pion wave function is identical to those
constructed from the pion transition and electromagnetic form
factors, consistent with its universality. Choosing the
renormalization and factorization scales
lower than the $B$ meson mass, the NLO corrections are under
control: they amount only up to 30\% of the form factors at large
recoil of the pion, when varying models for the meson wave
functions.

\end{abstract}


\maketitle

\section{INTRODUCTION}

$B$ meson transition form factors are an essential input of the
factorization approaches to nonleptonic two-body $B$ meson decays,
such as the perturbative QCD (PQCD) approach \cite{LY1,KLS} based
on the $k_T$ factorization theorem \cite{CCH,CE,LRS,BS,LS,HS}. For
next-to-leading-order (NLO) contributions in leading-twist PQCD,
the vertex corrections, the quark loops, and the magnetic penguins
associated with the weak decay vertices in factorizable emission
amplitudes have been calculated \cite{LMS05,LM06,LM062}. As
explained in \cite{LMS05}, the above corrections may be the most
crucial NLO pieces for understanding the known $B\to\pi\pi$ and
$B\to\pi K$ puzzles, which result from the large observed
$\pi^0\pi^0$ branching ratio, and from the dramatically different
direct CP asymmetries between the $\pi^\mp K^\pm$ and $\pi^0
K^\pm$ modes, respectively. There have been many applications of
this NLO PQCD formalism to nonleptonic two-body $B$ and $B_s$
meson decays in the literature. For NLO corrections to
spectator diagrams, we have identified the so-called Glauber
divergences, in additional to those which are absorbed into hadron
wave functions, and summed them into a phase factor to all orders
\cite{LM11}. It was observed that the phase factor modifies the
interference pattern between the spectator diagrams, and further
improves the resolution of the $B\to\pi\pi$, $\pi K$
puzzles in NLO PQCD. At the same level of accuracy, we need to
calculate NLO corrections to the $B$ meson transition form
factors for completeness.

In this paper we shall extend the NLO framework for the pion
electromagnetic form factor in the $k_T$ factorization \cite{LSW11} to the
$B\to\pi$ transition form factors. In this framework light partons
in both QCD quark diagrams and effective diagrams for hadron wave
functions are off mass shell by $k_T^2$ \cite{NL2,NL07}. Not only
the collinear divergences from gluon emissions collimated to the
pion, but also the soft divergences from gluon exchanges between
the two mesons exist. Compared to the pion form factor
\cite{LSW11}, a new point is that an infrared regulator
associated with the $b$ quark is not needed. Due to its finite mass,
gluons radiated by the $b$ quark do not generate collinear
divergences. Soft divergences can be regularized either by the
virtuality of internal particles, or by the virtuality $k_T^2$ of
other light partons, to which the radiative gluons attach. That is,
the $b$ quark remains on-shell in the above framework, a condition
which justifies the approximation of the $b$ quark field by the
effective heavy-quark field for defining the $B$ meson wave
function. The NLO pion wave function is found to be identical to those
constructed in the pion electromagnetic and transition form
factors \cite{LSW11,NL07}, consistent with its universality. Note that
the diagrams considered here differ from those in the QCD-improved
factorization (QCDF) approach \cite{BF01} and in the soft-collinear
effective theory (SCET) \cite{Bauer:2002aj}, which are based on the collinear
factorization theorem \cite{LB80}: there is no end-point singularity in the
$k_T$ factorization, so it is not necessary to introduce soft form
factors \cite{Beneke:2003pa} or to perform the zero-bin
subtraction \cite{MS06} in our calculation.

It will be demonstrated that the collinear and soft divergences in
the quark diagrams are cancelled by those in the effective diagrams
for the $B$ meson and pion wave functions. Taking the difference of
the above sets of diagrams, we derive the $k_T$-dependent NLO hard
kernel at leading twist for the $B\to\pi$ transition form factors.
The infrared finiteness of
the hard kernel confirms the application of the $k_T$ factorization
theorem to $B$ meson semileptonic decays \cite{TLS}. Similar to the analysis in
\cite{LSW11,NL07}, both the large double logarithms
$\alpha_s\ln^2k_T$ and $\alpha_s\ln^2 x$, $x$ being a parton
momentum fraction, are identified. The former is absorbed into the
$B$ meson and pion wave functions and summed to all orders in the coupling
constant $\alpha_s$ by the $k_T$ resummation \cite{LY1}, and the latter is
absorbed into a jet function and summed to all orders by the
threshold resummation \cite{UL}. Due to the dominant soft dynamics associated
with the $b$ quark, the effect of the $k_T$ resummation from the $B$
meson side is minor. The renormalization scale $\mu$ and the
factorization scale $\mu_{\rm f}$ are introduced by higher-order
corrections to the quark diagrams and to the effective diagrams,
respectively. Choosing $\mu$ and $\mu_{\rm f}$ appropriately, with
both being lower than the $B$ meson mass as postulated in
\cite{LS,KLS}, the NLO corrections are under control: they amount
only up to 30\% of the form factors at large recoil of the pion,
when varying models for the meson wave functions.

In Sec.~II we calculate the $O(\alpha_s^2)$ QCD quark diagrams for
the $B\to\pi\ell\bar\nu$ semileptonic decay, the $O(\alpha_s)$ effective
diagrams for the $B$ meson and pion wave functions, and their
convolutions with the $O(\alpha_s)$ hard kernel. Since the $k_T$
factorization is appropriate for QCD processes dominated by
contributions from small $x$ \cite{NL2}, we shall keep only terms in
leading power of $x$. The important double logarithms are
identified, and the $k_T$-dependent NLO hard kernel is presented. Section III
contains the numerical investigation, in which we examine the
dependence of the NLO contributions to the
$B\to\pi$ transition form factors on
the renormalization and factorization scales, and on the shape of
the $B$ meson and pion wave functions. Section IV is the conclusion.

\section{NLO CORRECTIONS}

\begin{figure}[t]
\begin{center}
\hspace{-3 cm}
\includegraphics[height=3 cm]{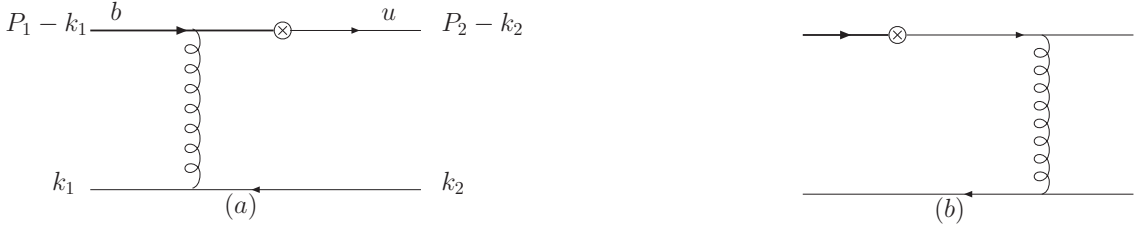}
\caption{Leading-order quark diagrams for the $B \to \pi$ transition form factors
with $\otimes$ representing the weak vertex.} \label{leading}
\end{center}
\end{figure}

In this section we calculate the $O(\alpha_s^2)$ quark diagrams for
the $B\to\pi \ell\bar\nu$
semileptonic decay, and the $O(\alpha_s)$ effective diagrams
for the $B$ meson and pion wave functions in the Feynman gauge.
The $B\to\pi$ transition form factors are defined via the matrix element
\begin{eqnarray}
\langle \pi (P_2) | \bar{u} \gamma^{\mu} b |  B(P_1)\rangle &=&
f^{+} (q^2) (P_1^{\mu}+P_2^{\mu}) +  [f^{0}
(q^2)-f^{+} (q^2)] {m_B^2 -m_{\pi}^2 \over q^2}
q^{\mu},\label{form}
\end{eqnarray}
where $m_B$ ($m_\pi$) is the $B$ meson (pion) mass,
and $q= P_1 -P_2$ is the transfer momentum.
The momentum $P_1$ ($P_2$) of the $B$ meson (pion) is chosen as
$P_1 =P_1^+(1,1,{\bf 0}_T)$ ($P_2 = (0,P_2^-,{\bf 0}_T)$) with the
component $P_1^+= m_B/\sqrt{2}$ ($P_2^-=\eta \, m_B/\sqrt{2}$).
The large recoil region of the pion corresponds to the energy
fraction $\eta\sim O(1)$. According to the $k_T$ factorization,
the anti-quark $\bar q$ carries the momentum $k_1=(x_1P_1^+,0,{\bf
k}_{1T})$ in the $B$ meson and $k_2=(0,x_2P_2^-,{\bf k}_{2T})$ in
the pion, $x_1$ and $x_2$ being the momentum fractions, as
labelled in the leading-order (LO) quark diagrams in Fig.~\ref{leading}. We
postulate the hierarchy
\begin{eqnarray}
m_B^2 \gg x_2 m_B^2 \gg x_1 m_B^2 \gg  x_1 x_2 m_B^2, \,
k_{1T}^2, \, k_{2T}^2, \label{hie}
\end{eqnarray}
in the small-$x$ region, which is roughly consistent with the
order of magnitude: $x_2\sim 0.3$, $x_1\sim 0.1$, $m_B\sim 5$ GeV,
and $k_T\lesssim 1$ GeV. Under the above hierarchy,
only those terms that do not vanish in the $x\to 0$ and $k_T
\to 0$ limits are kept, so the expressions of
our NLO results will be greatly simplified.

To obtain the LO hard kernels, we sandwich Fig.~\ref{leading} with
the following leading-twist spin projectors for the $B$ meson and
the pion \cite{BF01,HW05}
\begin{eqnarray}
\frac{1}{2\sqrt{N_c}}(\not\! P_1+m_B) \gamma_5\left[\not\!
n_+\phi_B^{(+)}(x_1) + \left (\not\! n_{-} - k_1^{+}
\gamma_{\perp}^{\nu} {
\partial \over \partial { {\bf k}^{\nu}_{1T}}} \right )
\phi_B^{(-)}(x_1)\right], \qquad  \frac{1}{\sqrt{2N_c}}\gamma_5
\not\! P_2 \phi_\pi(x_2) \,, \label{bwp2}
\end{eqnarray}
respectively, where the dimensionless vectors are defined by
$n_+=(1,0,{\bf 0}_T)$, and $n_-=(0,1,{\bf 0}_T)$ along $P_2$, and
$N_c$ is the number of colors. The contributions proportional to
the $B$ meson distribution amplitudes
$\phi_B^{(+)}$ and $\phi_B^{(-)}$ from Fig.~\ref{leading}(a) are
computed as
\begin{eqnarray}
H_a^{(0)}(x_1,k_{1T},x_2,k_{2T}) &=& -4g^2C_F \frac{[x_2 \eta
\phi_B^{(+)}(x_1) +\phi_B^{(-)}(x_1) ] {P_2}^{\mu}}{x_2 \eta (x_1
x_2 \eta m_B^2+|{\bf k}_{1T}-{\bf k}_{2T}|^2)} \phi_\pi(x_2),\label{fa0}
\end{eqnarray}
with the strong coupling $g$, and the color factor $C_F$. To reach
the above expression, we have applied the hierarchy $x_2m_B^2\gg
k_{2T}^2$ in Eq.~(\ref{hie}) to the internal $b$ quark propagator.
The denominator $x_1 x_2\eta m_B^2+|{\bf k}_{1T}-{\bf k}_{2T}|^2$
comes from the virtuality of the LO hard gluon, in which the
$|{\bf k}_{1T}-{\bf k}_{2T}|^2$ term smears the end-point singularity
from small $x_2$. Similarly, Fig.~\ref{leading}(b) leads to the amplitude
\begin{eqnarray}
H_b^{(0)}(x_1,k_{1T},x_2,k_{2T})
&=& -4g^2C_F \frac{(\eta {P_1}^{\mu}- {P_2}^{\mu})\phi_B^{(+)}
(x_1)+ {P_2}^{\mu} \phi_B^{(-)}(x_1) }{\eta (x_1 x_2 \eta
m_B^2+|{\bf k}_{1T}-{\bf k}_{2T}|^2)} \phi_\pi(x_2). \label{fd0}
\end{eqnarray}
Comparing Eqs.~(\ref{fa0}) and (\ref{fd0}), it is easy to see that 
the term proportional to $\phi_B^{(-)}$ from Fig.~\ref{leading}(a) 
dominates numerically according to the hierarchy in Eq.~(\ref{hie}). 
As explained above, the $B\to\pi$ form factors receive major 
contributions from the small-$x$ region, in which the $k_T$ factorization 
is an appropriate framework. Since the amplitude from Fig.~\ref{leading}(b) 
is suppressed by a power of $x_2$, we will not consider the NLO 
corrections to $H_b^{(0)}(x_1,k_{1T},x_2,k_{2T})$, and focus on those 
to Fig.~\ref{leading}(a) below. The term proportional to ${P_1}^{\mu}$ 
in Eq.~(\ref{fd0}) gives the symmetry breaking effect
\cite{BF01}, which is calculable even in the collinear
factorization, as convoluted with $\phi_B^{(+)}(x_1) \sim x_1$ at
small $x_1$.

\subsection{NLO Quark Diagrams}

\begin{figure}[t]
\begin{center}
\hspace{-2 cm}
\includegraphics[height=8cm]{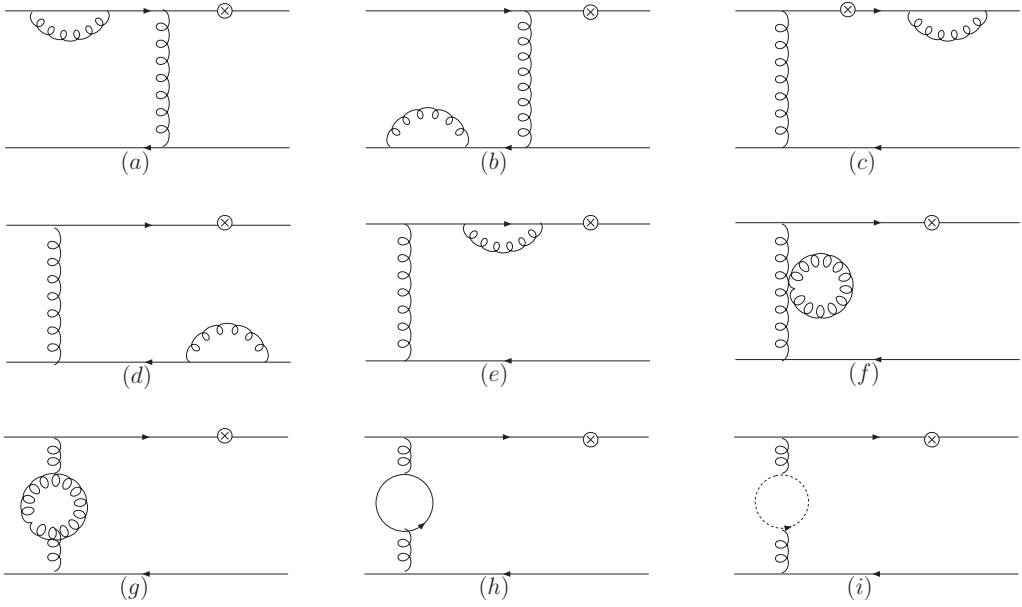}
\caption{Self-energy corrections to Fig.~\ref{leading}(a).}
\label{selfd B}
\end{center}
\end{figure}

\begin{figure}[t]
\begin{center}
\hspace{-2 cm}
\includegraphics[height=6cm]{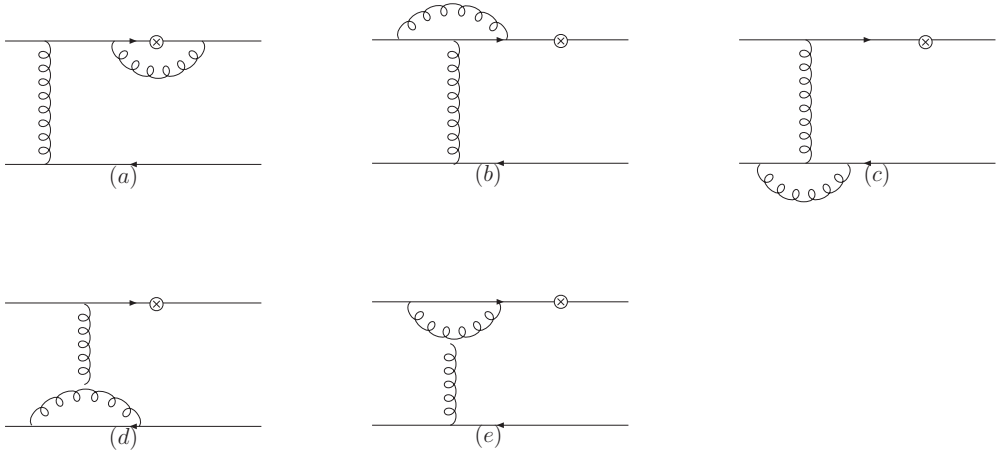}
\caption{Vertex corrections to Fig.~\ref{leading}(a).}
\label{three B}
\end{center}
\end{figure}

\begin{figure}[t]
\begin{center}
\hspace{-3 cm}
\includegraphics[height=5 cm]{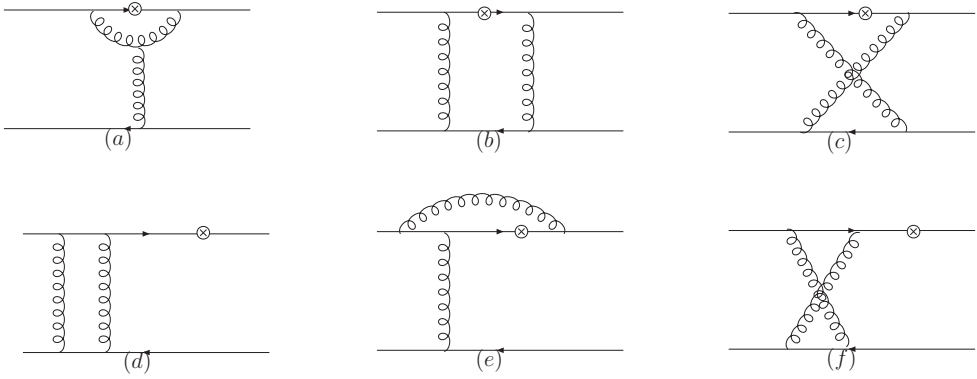}
\caption{Box and pentagon diagrams.} \label{four}
\end{center}
\end{figure}

The NLO corrections to Fig.~\ref{leading}(a) contain
Figs.~\ref{selfd B}, \ref{three B}, and \ref{four} for the
self-energy corrections, the vertex corrections, and the box and
pentagon diagrams, respectively. The ultraviolet poles are
extracted in the dimensional reduction \cite{WS79} in order to
avoid the ambiguity from handling the matrix $\gamma_5$.
We adopt the following convenient dimensionless ratios
\begin{eqnarray}
& &\delta_{1} =\frac{k_{1T}^2}{m_B^2},\;\;\;\; \delta_{2}
=\frac{k_{2T}^2}{m_B^2},\nonumber\\
& &\delta_{12} =\frac{x_1 x_2 \eta m_B^2 + {|{\bf k}_{1T}-{\bf
k}_{2T}|}^2}{m_B^2},
\end{eqnarray}
as presenting our results. The infrared poles are then
identified as the logarithms $\ln\delta_1$ and $\ln\delta_2$.

The self-energy corrections in Fig.~\ref{selfd B} give
\begin{eqnarray}
G^{(1)}_{2a} &=& - \frac{\alpha_s C_F}{4 \pi} \left[ {6 \over
\delta_1} \bigg (\frac{1}{\epsilon}+\ln\frac{4\pi\mu^2 }{m_B ^2
e^{\gamma_E}}  + {5 \over 3} \bigg)+ {1 \over 2}  \left({1 \over
\epsilon} +\ln\frac{4\pi\mu^2 }{m_B ^2 e^{\gamma_E}} + 2\ln {m_g^2
\over m_B^2 }  - 1 \right) \right ] H^{(0)},\label{g2a}
\\
G^{(1)}_{2b} &=& -\frac{\alpha_s C_F}{8\pi}
\left[\frac{1}{\epsilon} +\ln\frac{4\pi\mu^2}{\delta_1 m_B^2
e^{\gamma_E}}+2\right] H^{(0)},
\\
G^{(1)}_{2c,2d}&=&-\frac{\alpha_sC_F}{8\pi}
\left[\frac{1}{\epsilon} +\ln\frac{4\pi\mu^2}{\delta_2 m_B^2
e^{\gamma_E}}+2\right] H^{(0)},
\\
G^{(1)}_{2e}&=& - \frac{\alpha_s C_F}{4 \pi}   \left[
{6 \over x_2 \eta} \bigg (\frac{1}{\epsilon}+\ln\frac{4\pi\mu^2
}{m_B ^2 e^{\gamma_E}} + {5 \over 3} \bigg)+ \left({1 \over
\epsilon} +\ln\frac{4\pi\mu^2 }{m_B ^2 e^{\gamma_E}} + 4 \ln (x_2
\eta) -5 \right) \right ] H^{(0)},
\\
G^{(1)}_{2f+2g+2h+2i}&=&\frac{\alpha_s}{4\pi} \left [
\left(\frac{5}{3}N_c-\frac{2}{3}N_f\right)\left(
\frac{1}{\epsilon}+\ln\frac{4\pi\mu^2 }{\delta_{12}
m_B^2e^{\gamma_E}}\right) \right ] H^{(0)},
\end{eqnarray}
where $1/\epsilon$ represents the ultraviolet pole, $\mu$ is the
renormalization scale,  $\gamma_E$ is the Euler constant, $N_f$ is
the number of quark flavors, and $H^{(0)}$ denotes the
leading-twist LO hard kernel proportional to ${P_2}^{\mu}$,
\begin{eqnarray}
H^{(0)}(x_1,k_{1T},x_2,k_{2T})&=& - \frac{4g^2C_F {P_2}^{\mu}}{x_2
\eta \delta_{12} m_B^2 }.\label{H0}
\end{eqnarray}

The above expressions are basically similar to the corresponding ones
obtained in the pion electromagnetic form factor \cite{LSW11}. We
emphasize only that Fig.~\ref{selfd B}(a), the self-energy correction
to the $b$ quark, requires a mass renormalization as indicated by the
first term in the square brackets of Eq.~(\ref{g2a}).
The finite piece of the first term is then absorbed, with the relation
$(P_1-k_1)^2-m_b^2=-k_{1T}^2$, into the redefinition the $b$ quark mass,
\begin{eqnarray}
\frac{1}{(P_1-k_1)^2-m_b^2}\left[1- \frac{\alpha_s C_F}{4 \pi} {6
\over \delta_1} \bigg (\ln\frac{\mu^2 }{m_B ^2}  + {5\over 3}
\bigg)\right]= \frac{1}{(P_1-k_1)^2-m_b^2(\mu)},
\end{eqnarray}
leading to the pole mass
\begin{eqnarray}
m_b(\mu)=m_b\left[1+\frac{\alpha_s}{\pi}
\bigg (\ln\frac{\mu^2 }{m_B ^2}  + {5\over 3} \bigg)\right].
\end{eqnarray}
In this work we shall not differentiate $m_b(\mu)$ from $m_B$,
because the distinction between them contributes at
next-to-leading power. The second term in the square brackets of
Eq.~(\ref{g2a}) represents the correction to the $b$ quark wave
function. As explained before, we shall consider an on-shell
valence $b$ quark, so the involved soft divergence is regularized
by a gluon mass $m_g$, which will be cancelled by the
corresponding soft divergence in the effective diagram
Fig.~\ref{ktlc2 B}(a) below.

The results from the vertex corrections in Fig.~\ref{three B}
are summarized as
\begin{eqnarray}
G_{3a}^{(1)} &=& { \alpha_s  C_F  \over 4 \pi}   \bigg
[{1 \over \epsilon}  + \ln { 4 \pi \mu^2 \over m_B^2 e^{\gamma_E}}
- 2 \ln \bigg({\delta_2 \over \eta}\bigg) (1 + \ln x_2) + \ln^2
x_2  - { \pi^2 - 3\over 2} \bigg ] H^{(0)}   ,
\\ G_{3b}^{(1)}&=&   -{ \alpha_s   \over 8 \pi N_c}
\bigg [ {1 \over \epsilon}  + \ln { 4 \pi \mu^2 \over m_B^2
e^{\gamma_E} } + 4 \ln ({x_2 \eta}) \bigg ] H^{(0)} ,
\label{3b}\\
G_{3c}^{(1)} &=& -{ \alpha_s   \over 8 \pi N_c}
\bigg [ \frac{1}{\epsilon}+\ln\frac{4\pi\mu^2}{\delta_{12}
m_B^2e^{\gamma_E}}   -  2 \ln \bigg({ \delta_{1}\over
\delta_{12}}\bigg )\ln \bigg ({ \delta_{2}\over \delta_{12}}
\bigg) - 2 \ln \bigg ({\delta_1 \delta_2 \over \delta^2_{12}}
\bigg) - {2 \pi^2 \over 3} + {9 \over 2} \bigg ] H^{(0)},
\label{3c}\\
G_{3d}^{(1)}&=& \frac{\alpha_s N_c}{8 \pi} \bigg
[\frac{3}{\epsilon} - 3 \gamma_E  + 3 \ln { 4 \pi \mu^2 \over
\delta_{12} m_B^2 e^{\gamma_E} } + 2 \ln \bigg ({ \delta_{12}^2
\over \delta_1 \delta_2 } \bigg ) + 7 \bigg ] H^{(0)} ,
\label{3d} \\
G_{3e}^{(1)}&=&  \frac{\alpha_s N_c}{8 \pi}   \bigg [
\frac{3}{\epsilon}+ 3 \ln\frac{4\pi\mu^2}{ m_B^2e^{\gamma_E}}- {1
\over 2} \ln^2 \bigg ({ \delta_{12} \over \eta^2} \bigg ) + 2 (\ln
x_2-1) \ln \bigg({x_1 \over \eta}\bigg) - { \pi^2 \over 2} +3
\bigg ] H^{(0)}. \label{3e}
\end{eqnarray}
The amplitude from Fig.~\ref{three B}(a) depends only on the
regulator $\delta_2$, because the radiative gluon attaches to the
virtual $b$ quark line. The double logarithm $2\ln \delta_2 \ln x_2$
leads to the known Sudakov logarithm $\ln^2\delta_2$ and the known
threshold logarithm $\ln^2x_2$ \cite{NL07,LSW11}, as reexpressed
in the form
\begin{eqnarray}
2\ln\delta_2\ln x_2=\ln^2\delta_2+\ln^2x_2
-\ln^2\frac{\delta_2}{x_2}.\label{dou}
\end{eqnarray}
The radiative gluon in Fig.~\ref{three B}(b) attaches to the
massive valence $b$ quark and the virtual $b$ quark, so Eq.~(\ref{3b})
is infrared finite. The radiative gluon in Fig.~\ref{three B}(c)
attaches to the light valence anti-quarks, such that both the collinear and soft
divergences are produced, with the latter being denoted by the product
$\ln\delta_1\ln\delta_2$. This term can be absorbed neither into the
$B$ meson wave function nor into the pion wave function.
Since the radiative gluon attaches to the virtual LO hard gluon
in Fig.~\ref{three B}(d), the soft divergence does not appear
Eq.~(\ref{3d}). Equations~(\ref{3c}) and (\ref{3d}) are
symmetric under the exchange of the regulators $\delta_1$ and
$\delta_2$, as they should. Similar to Fig.~\ref{three B}(b),
Fig.~\ref{three B}(e) also gives an infrared finite contribution.

The box diagrams and the pentagon diagrams in Fig.~\ref{four}
lead to the amplitudes
\begin{eqnarray}
G^{(1)}_{4a}&=&  - \frac{\alpha_s  N_c }  {4 \pi}  \bigg
[\ln \bigg ({  x_2 \eta^2 \over \delta_2 }\bigg )+ 1  \bigg ] x_2H^{(0)},
\label{4a}\\
G_{4c}^{(1)}&=& - \frac{\alpha_s}  {4 \pi N_c}
\bigg [  \ln \bigg ({ x_1 \eta \over \delta_{1}} \bigg ) \ln \bigg
({ \delta_{12} \over \delta_2}\bigg ) + { \pi^2 \over 6} \bigg ]
H^{(0)},
\\
G_{4d}^{(1)}&=& - \frac{\alpha_s C_F}  {4 \pi} \bigg [ \ln^2
\bigg({\delta_{1} \over x_1^2 }\bigg)  -  \ln^2 {x_1} -{7 \pi^2
\over 3} \bigg ] H^{(0)},
\\
G_{4e}^{(1)}&=& \frac{\alpha_s}  {8 \pi  N_c}   \bigg
[ \ln^2 \bigg({x_2 \eta^2  \over \delta_2 }\bigg) + \pi^2 \bigg ]
H^{(0)} ,
\label{4e}\\
G_{4f}^{(1)}&=& \frac{\alpha_s}  {8 \pi  N_c}  \bigg
[   \ln \bigg({\delta_{12} \over \delta_2 }\bigg) \bigg ( \ln
({\delta_{12}  \delta_2 }) - 4 \ln ({x_2 \eta }) \bigg ) \bigg ]
H^{(0)}.\label{4f}
\end{eqnarray}
Note that Eq.~(\ref{4a}) is power-suppressed in the small $x_2$
region, while the corresponding diagram gives a leading amplitude
in the pion form factor \cite{LSW11}. The difference is attributed
to the spin projectors: it is $\not\!n_-\propto \gamma^+$ on the
$B$ meson side here, but $\not\!P_1\propto \gamma^-$ on the
initial pion side in the latter case. Simply counting the sequence
of the gamma matrices, it is easy to understand that
Fig.~\ref{four}(a) does not produce an amplitude proportional to
$H^{(0)}$ at leading power. Figure~\ref{four}(b) is a two-particle
reducible diagram, so its contribution will be cancelled by the
corresponding effective diagram for the pion wave function
\cite{LSW11}, and needs not to be computed. Figure~\ref{four}(c)
also contains the soft divergence denoted by the
$\ln\delta_1\ln\delta_2$ term, which cancels that in
Fig.~\ref{three B}(c). It seems that Fig.~\ref{four}(d) generates
a collinear divergence, as the gluon on the right is parallel to
the light anti-quark in the pion. However, a careful look at the
sequence of the gamma matrices, similar to that for
Fig.~\ref{four}(a), reveals power suppression on this collinear
divergence. Equation~(\ref{4e}) does not depend on an infrared
regulator associated with the massive valence $b$ quark, because
$\delta_2$ alone is enough to regularize the collinear and soft
divergences. The collinear divergence associated with the light
valence anti-quark on the $B$ meson side is also power-suppressed
in Fig.~\ref{four}(f), so Eq.~(\ref{4f}) does not contain
$\ln\delta_1$.

The amplitudes from all the NLO quark diagrams are summed into
\begin{eqnarray}
G^{(1)} &=& { \alpha_s C_F \over 4 \pi}   \bigg [ {21 \over 4}
\left (  {1 \over \epsilon} + \ln { 4 \pi \mu^2 \over m_B^2
e^{\gamma_E}}\right ) -\ln^2 \delta_1 + \left(4 \ln x_1 -{3 \over
2}\right)  \ln \delta_1  + \ln {m_B^2 \over m_g^2} - \left( 2 \ln
x_2 + 3 \right ) \ln \delta_2    \nonumber \\
&& -{55 \over 16} \ln^2 x_1 + {7 \over 16} \ln^2 x_2 + { 9 \over
8} \ln x_1 \ln x_2+ { 7 \ln \eta -18 \over 8} \ln x_1 + { 7 \ln
\eta -36  \over 8} \ln x_2 \nonumber \\
&& -{ \ln \eta (7 \ln \eta +4 )\over 16} + { 23 \over 16 } \pi^2
+{235 \over 16} \bigg] H^{(0)}, \label{pgt}
\end{eqnarray}
for $N_f=6$. The ultraviolet pole in the above expression is
the same as in the pion electromagnetic form factor, which determines the
renormalization-group (RG) evolution of the coupling constant
$\alpha_s$.

\subsection{NLO Effective Diagrams}

The $O(\alpha_s)$ $B$ meson wave function $\Phi_{B}^{(1)}$
\cite{Li:2004ja} and the $O(\alpha_s)$ pion wave function
$\Phi_{\pi}^{(1)}$ \cite{NL2,L1} collect the effective diagrams from
the matrix elements of the leading Fock states
\begin{eqnarray}
\Phi_{B}(x_1,k_{1T};x'_1,k'_{1T})&=&\int\frac{dz^-}{2\pi
}\frac{d^2z_T}{(2\pi)^2}e^{-ix'_1P_1^+ z^-+i{\bf k}'_{1T}\cdot
{\bf z}_T}  \langle 0|{\bar q}(z) W_z(n_1)^{\dag} W_0(n_1)
\not n_-\Gamma h_v(0)|h_v\bar q(k_1)\rangle,\label{de2}\\
\Phi_{\pi}(x_2,k_{2T};x'_2,k'_{2T})&=&\int\frac{dy^+}{2\pi
}\frac{d^2y_T}{(2\pi)^2}e^{-ix'_2P_2^- y^+ +i{\bf k}'_{2T}\cdot
{\bf y}_T}\nonumber\\
& &\times\langle 0|{\bar q}(y) W_y(n_2)^{\dag} W_0(n_2) \not
n_{+}\gamma_5 q(0)|u(P_2-k_2)\bar q(k_2)\rangle,\label{de1}
\end{eqnarray}
respectively, with $z=(0,z^-,{\bf z}_T)$ and
$y=(y^+,0,{\bf y}_T)$ being the coordinates of the anti-quark field
$\bar q$, respectively, $h_v$ the effective heavy-quark field,
and $\Gamma$ an appropriate gamma matrix.
In the above expressions the Wilson line $W_z(n_1)$ with $n_1^2\not=0$ is
written as
\begin{eqnarray}
\label{eq:WL.def} W_z(n_1) = P \exp\left[-ig \int_0^\infty d\lambda
n_1\cdot A(z+\lambda n_1)\right],
\end{eqnarray}
and the definition of the Wilson line $W_y(n_2)$ is similar. It is
understood that the
two Wilson lines $W_z(n_1)$ and $W_0(n_1)$ ($W_y(n_2)$ and
$W_0(n_2)$) are connected by a vertical link at infinity
\cite{BJY,CS08}. Equation~(\ref{de2}) ((\ref{de1})) produces
additional light-cone singularities \cite{Co03,Li:2004ja,MW}
from the region with a loop momentum collinear to $n_-$ ($n_+$), as the
Wilson line direction approaches the light cone, i.e., as $n_1\to
n_-$ ($n_2\to n_+$) \cite{Co03}. Hence, $n_1^2$ and $n_2^2$ serve
as the infrared regulators for the light-cone singularities in our
formalism. The $B$ meson and pion wave functions
then depend on the scales $\zeta_1^2\equiv 4(n_1 \cdot
P_1)^2/|n_1^2|$ and $\zeta_2^2\equiv 4(n_2 \cdot P_2)^2/|n_2^2|$,
respectively, whose variation is regarded as a factorization-scheme
dependence. This scheme dependence, entering the hard
kernel when taking the difference between the quark diagrams and
the effective diagrams, can be minimized by adhering to fixed
$n_1^2$ and $n_2^2$.

\begin{figure}[t]
\begin{center}
\hspace{-2.5 cm}
\includegraphics[height=6 cm]{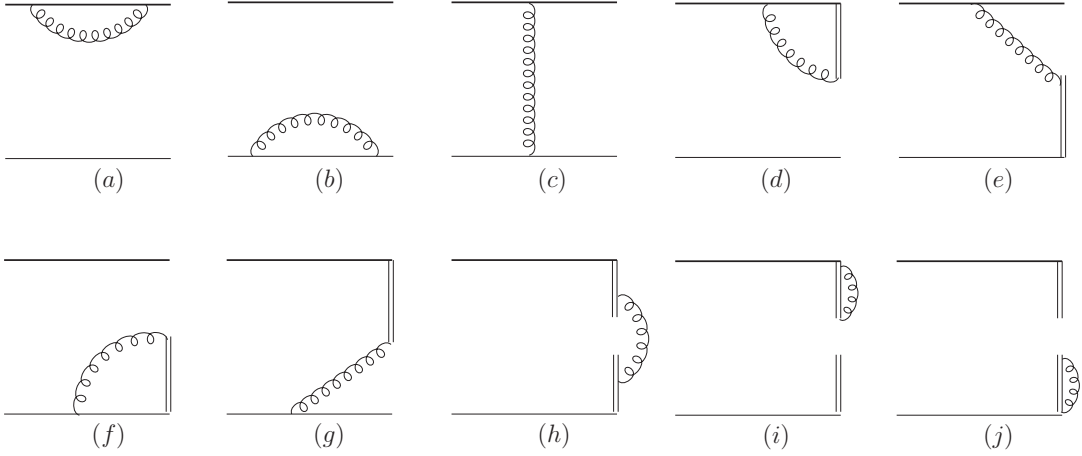}
\caption{$O(\alpha_s)$ diagrams for the $B$ meson wave function.} \label{ktlc2 B}
\end{center}
\end{figure}

We compute the convolution of the NLO wave function
$\Phi_B^{(1)}$ with the LO hard kernel $H^{(0)}$ over the
integration variables $x_1'$ and ${\bf k'}_{1T}$,
\begin{eqnarray}
\Phi^{(1)}_{B}\otimes H^{(0)}&\equiv& \int dx'_1 d^2 {\bf k'}_{1T}
\Phi^{(1)}_{B}(x_1, {\bf k}_{1T}; x'_1,{\bf k'}_{1T})H^{(0)}(x'_1,
{\bf k'}_{1T} ,x_2, {\bf k}_{2T}). \label{Bdef}
\end{eqnarray}
The sign of the plus component $n_1^+$ of the vector $n_1$ is
arbitrary, which could be positive or negative ($n_1^-$ has a
positive sign, the same as of $P_2^-$). Choosing $n_1^+<0$, i.e.,
$n_1^2<0$ as in \cite{LS,LY1,Li96}, we derive, from
Figs.~\ref{ktlc2 B}(a)-\ref{ktlc2 B}(g),
\begin{eqnarray}
\Phi^{(1)}_{5a} \otimes H^{(0)} &=&  \frac{\alpha_s C_F}{4 \pi}
\left(\frac{1}{\epsilon}+\ln\frac{4\pi\mu_{\rm f}^2 }{m_g ^2
e^{\gamma_E}}
\right) H^{(0)}, \\
\Phi^{(1)}_{5b} \otimes H^{(0)} &=& -\frac{\alpha_s
C_F}{8\pi}\left(\frac{1}{\epsilon} +\ln\frac{4\pi\mu_{\rm
f}^2}{\delta_1
m_B^2 e^{\gamma_E}}+2\right) H^{(0)}, \\
\Phi^{(1)}_{5c} \otimes H^{(0)} &=& -\frac{\alpha_s
C_F}{4 \pi} \left( \ln^2 {\delta_1 \over x_1^2} \right) H^{(0)}, \\
\Phi^{(1)}_{5d} \otimes H^{(0)} &=& - \frac{\alpha_s C_F}{4 \pi}
\ln {\zeta_1^2 \over m_B^2}
\left(\frac{1}{\epsilon}+\ln\frac{4\pi\mu_{\rm f}^2 }{m_g ^2
e^{\gamma_E}}
\right) H^{(0)}, \label{bwd}\\
\Phi^{(1)}_{5e} \otimes H^{(0)} &=&  \frac{\alpha_s C_F}{4 \pi}
\ln {\zeta_1^2 \over m_B^2} \left(\ln\frac{\zeta_1^2 }{m_g^2} + {1
\over 2} \ln {\zeta_1^2 \over m_B^2}+ 2 \ln x_1
\right) H^{(0)}, \label{bwe}\\
\Phi^{(1)}_{5f} \otimes H^{(0)} &=& \frac{\alpha_s C_F}{4 \pi}
\left( {1 \over \epsilon}+ \ln\frac{4\pi\mu_{\rm f}^2 }{x_1^2
\zeta_1^2 e^{\gamma_E}} - \ln^2 { \delta_1 m_B^2 \over x_1^2
\zeta_1^2 }- 2 \ln { \delta_1 m_B^2 \over x_1^2 \zeta_1^2} +
{\pi^2 \over 3}
\right) H^{(0)}, \label{bwf}\\
\Phi^{(1)}_{5g} \otimes H^{(0)} &=& \frac{\alpha_s C_F}{4 \pi}
\left( \ln^2 { \delta_1 m_B^2 \over x_1^2 \zeta_1^2} - {2  \pi^2
\over 3} \right) H^{(0)},\label{bwg}
\end{eqnarray}
$\mu_{\rm f}$ being the factorization scale.
The two-particle reducible diagrams Figs.~\ref{ktlc2
B}(a) and \ref{ktlc2 B}(c) are calculated, since the effective
heavy-quark field employed in the $B$ meson wave function differs
from the $b$ quark field in the quark diagrams. Though the effective
diagrams and the quark diagrams have
the same soft poles, the finite pieces are different, which
contribute to the NLO hard kernel. The self-energy corrections to
the Wilson lines in Figs.~\ref{ktlc2 B}(h)-\ref{ktlc2 B}(j) yield
\begin{eqnarray}
\bigg(\Phi^{(1)}_{5h}+\Phi^{(1)}_{5i}+B^{(1)}_{5j}\bigg) \otimes
H^{(0)} &=&\frac{\alpha_sC_F}{2\pi} \bigg(\frac{1}{\epsilon}
+\ln\frac{4\pi\mu_{\rm f}^2}{\delta_{12} m_B^2e^{\gamma_E}}\bigg
)H^{(0)},\label{bwil}
\end{eqnarray}
the same as in the pion wave function \cite{LSW11}.

It is pointed out that the gluon mass $m_g$ has been adopted to
regularize the soft divergences in the diagrams involving the
effective heavy-quark field, namely, Figs.~\ref{ktlc2 B}(a),
\ref{ktlc2 B}(d), and \ref{ktlc2 B}(e). The soft divergence in
Fig.~\ref{ktlc2 B}(a) indeed cancels that in Fig.~\ref{selfd B}(a)
as stated in the previous subsection. The $m_g$ dependence
disappears in the sum of Eqs.~(\ref{bwd}) and (\ref{bwe}), which
must be the case, because the gluons emitted by the $b$ quark and
attaching to other particle lines do not generate soft
divergences. The hierarchy $\zeta_1^2\gg m_B^2$ was employed in
the derivation of Eq.~(\ref{bwe}) \cite{Li:2004ja}, so the large
double logarithm $\ln^2(\zeta_1^2/m_B^2)$ demands an additional
resummation treatment of the $B$ meson wave function, which will
not be performed in this work. The double logarithms
$\ln^2\delta_1$ from the quark diagram Fig.~\ref{four}(d) and from
the effective diagram Fig.~\ref{ktlc2 B}(c) cancel each other. The
double logarithms $\ln^2 \left(m_B^2 \delta_1  / (x_1^2 \zeta_1^2)
\right)$ are not only attenuated by $x_1^2$, but also cancel
exactly between Eqs.~(\ref{bwf}) and (\ref{bwg}). Summing all the
above $O(\alpha_s)$ contributions, we obtain
\begin{eqnarray}
\Phi^{(1)}_{B} \otimes H^{(0)} &=& { \alpha_s C_F \over 4 \pi}
\bigg [  \left (  \ln {m_B^2 \over \zeta_1^2} + {7 \over 2}\right
) \left ( {1 \over \epsilon} + \ln { 4 \pi \mu_{\rm f}^2 \over
m_B^2 e^{\gamma_E}}\right )  -\ln^2 \delta_1 +   \left(4 \ln x_1 -
{3 \over 2} \right) \ln \delta_1 + \ln {m_B^2 \over m_g^2}  \nonumber \\
&& + {3 \over 2} \ln^2 { m_B^2 \over \zeta_1^2} - ( 2 \ln x_1 -1)
\ln { m_B^2 \over \zeta_1^2} - 4 \ln^2 x_1 - 2 \ln ({x_2 \eta}) -{
\pi^2 \over 3} -1 \bigg ] H^{(0)}.\label{bpt}
\end{eqnarray}

\begin{figure}[t]
\begin{center}
\hspace{-2.5 cm}
\includegraphics[height=6.0 cm]{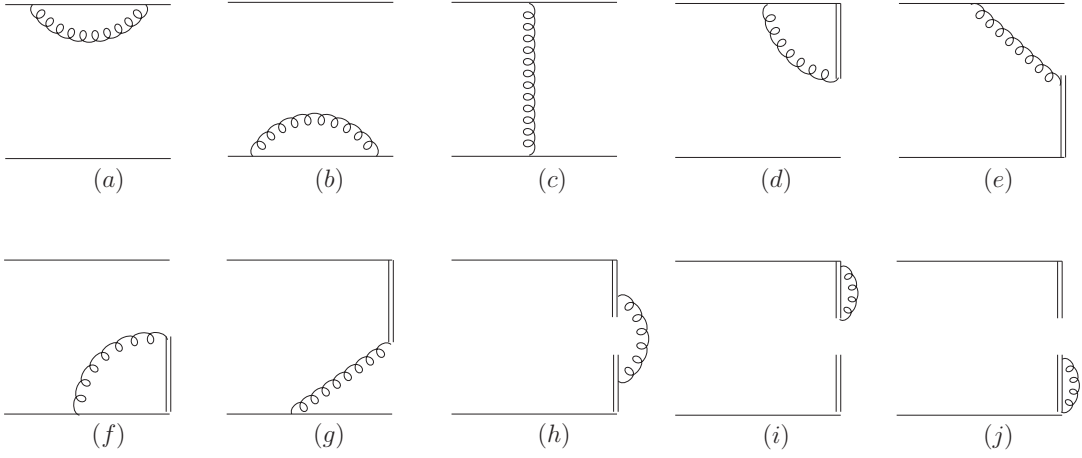}
\caption{$O(\alpha_s)$ diagrams for the pion wave
function.} \label{ktlc2 pi}
\end{center}
\end{figure}

We then compute the convolution of the NLO wave function
$\Phi_\pi^{(1)}$ with the LO hard kernel $H^{(0)}$ over the
integration variables $x_2'$ and ${\bf k'}_{2T}$,
\begin{eqnarray}
H^{(0)}\otimes \Phi^{(1)}_{\pi}&\equiv& \int dx'_2 d^2 {\bf
k'}_{2T} H^{(0)} (x_1,{\bf k}_{1T},x'_2,{\bf k'}_{2T})
\Phi^{(1)}_{\pi}(x_2,{\bf k}_{2T};x'_2,{\bf k'}_{2T}).
\end{eqnarray}
The corrections from Figs.~\ref{ktlc2 pi}(a)-\ref{ktlc2 pi}(j) are summarized as
\begin{eqnarray}
H^{(0)}\otimes\Phi^{(1)}_{6 a}&=& H^{(0)}\otimes\Phi^{(1)}_{6
b}=-\frac{\alpha_s C_F}{8\pi}\left(\frac{1}{\epsilon}
+\ln\frac{4\pi\mu_{\rm
f}^2}{\delta_2 m_B^2e^{\gamma_E}}+2\right)H^{(0)}, \label{pwb}\\
H^{(0)}\otimes\Phi^{(1)}_{6 c}&=&0, \\
H^{(0)}\otimes\Phi^{(1)}_{6 d}&=&
\frac{\alpha_sC_F}{4\pi}\left(\frac{1}{\epsilon}
+\ln\frac{4\pi\mu_{\rm f}^2}{\delta_2 m_B^2 e^{\gamma_E}}
-\ln^2\frac{\zeta_2^2}{\delta_2
m_B^2}+\ln\frac{\zeta_2^2}{\delta_2 m_B^2}+2-\frac{\pi^2}{3}
\right) H^{(0)}, \label{pwd}\\
H^{(0)}\otimes \Phi^{(1)}_{6 e}&=&
\frac{\alpha_sC_F}{4\pi}\left(\ln^2\frac{x_2  \zeta_2^2} {
\delta_2 m_B^2} + \pi^2 \right)H^{(0)}, \label{pwe}\\
H_b^{(0)}\otimes\Phi^{(1)}_{6 f}&=&
\frac{\alpha_sC_F}{4\pi}\left(\frac{1}{\epsilon}
+\ln\frac{4\pi\mu_{\rm f}^2}{\delta_2 m_B^2 e^{\gamma_E}}
-\ln^2\frac{x_2^2\zeta_2^2}{\delta_2 m_B^2}
+\ln\frac{x_2^2\zeta_2^2}{\delta_2 m_B^2}+2-\frac{\pi^2}{3}
\right)H^{(0)},\label{pwf}\\
H^{(0)}\otimes\Phi^{(1)}_{6 g}&=&  \frac{\alpha_sC_F}{4 \pi}
\left(\ln^2 \frac{ \delta_2m_B^2} {x_2^2   \zeta_2^2}   -  {\pi^2
\over 3}  \right)H^{(0)}, \label{pwg}\\
H^{(0)}\otimes\bigg(\Phi^{(1)}_{6 h}+\Phi^{(1)}_{6
i}+\Phi^{(1)}_{6 j}\bigg) &=&\frac{\alpha_sC_F}{2\pi}
\bigg(\frac{1}{\epsilon} +\ln\frac{4\pi\mu_{\rm f}^2}{\delta_{12}
m_B^2e^{\gamma_E}}\bigg )H^{(0)}, \label{pwh}
\end{eqnarray}
which are similar to those extracted from the pion transition
and electromagnetic form factors \cite{NL07,LSW11}, but with the hard
scale $Q$ being replaced by $m_B$ here. This similarity supports the
universality of the pion wave function.
Summing all the above $O(\alpha_s)$ contributions, we have
\begin{eqnarray}
H^{(0)}\otimes \Phi^{(1)}_{\pi}&=& { \alpha_s C_F \over 4 \pi}
\bigg [  3 \left (  {1 \over \epsilon} + \ln { 4 \pi \mu_{\rm f}^2 \over
m_B^2 e^{\gamma_E}}\right ) - \ln \delta_2  (2 \ln x_2 +3) + 2 \ln
{ \zeta_2^2 \over m_B^2} (\ln x_2 +1)  \nonumber \\
&& - 2 \ln \delta_{12} + \ln x_2 (\ln x_2 +2) + 2 \bigg ]
H^{(0)}. \label{ppt}
\end{eqnarray}

We stress that the ultraviolet poles are different in
Eqs.~(\ref{bpt}) and (\ref{ppt}), since the former involves the
effective heavy-quark field, instead of the $b$ quark field. That is,
the $B$ meson and pion wave functions exhibit different evolutions
as illustrated below. First, the $B$ meson decay constant,
defined via the matrix element with the effective heavy-quark field,
evolves with an energy scale. Hence, part of the $\ln\mu_{\rm f}$
term in Eq.~(\ref{bpt}) should be absorbed into $f_B(\mu_{\rm f})$
through the RG equation in the heavy quark effective theory (HQET)
\begin{eqnarray}
\left(\mu {d \over d \mu} + { \alpha_s  C_F \over 4 \pi } \gamma_f
\right) f_B(\mu_{\rm f})=0,
\end{eqnarray}
with the anomalous dimension $\gamma_f=-3$ at one loop
\cite{fB evolution}. The RG equation for the $B$ meson wave function
without the decay constant,
$\Phi_{B}(x_1, \mu_f) /f_B(\mu_{\rm f})$, is then written as
\begin{eqnarray}
\left(\mu {d \over d \mu} + { \alpha_s  C_F \over 4 \pi } \gamma_B
\right) {\Phi_{B}(x_1, \mu_f) \over  f_B(\mu_{\rm f})} =0,
\end{eqnarray}
where the anomalous dimension
\begin{eqnarray}
\gamma_B = -2 \, \left (  \ln {m_B^2 \over \zeta_1^2} + 2 \right ),
\end{eqnarray}
governs part of the RG
evolution in the $k_T$ factorization formulas for the $B\to\pi$
form factors \cite{LY1,TLS}.

\subsection{NLO Hard Kernel}

The infrared-finite $k_T$-dependent NLO hard kernel for the $B\to\pi$ transition
form factors is derived by taking the difference between the
quark diagrams and the effective diagrams \cite{NL2}
\begin{eqnarray}
H^{(1)}(x_1,{\bf k}_{1T},x_2,{\bf k}_{2T})&=&G^{(1)}(x_1,{\bf k}_{1T},x_2,{\bf k}_{2T})\nonumber\\
& &-\int dx'_1d^2{\bf k'}_{1T} \Phi^{(1)}_{B}(x_1,{\bf k}_{1T};
x'_1,{\bf k'}_{1T}) H^{(0)}(x'_1,{\bf k'}_{1T},x_2,{\bf k}_{2T})\nonumber \\
& &-\int dx'_2 d^2 {\bf k'}_{2T}H^{(0)}(x_1,k_{1T},x'_2,{\bf
k'}_{2T}) \Phi^{(1)}_{\pi}(x_2,{\bf k}_{2T};x'_2,{\bf k'}_{2T}).
\label{pa1}
\end{eqnarray}
Note that $\alpha_s$ appearing in
Eqs.~(\ref{pgt}), (\ref{bpt}), and (\ref{ppt}) denotes the bare
coupling constant, which can be rewritten as
\begin{eqnarray}
\alpha_s=\alpha_s(\mu_{\rm f})+\delta Z(\mu_{\rm
f})\alpha_s(\mu_{\rm f}),\label{dz}
\end{eqnarray}
with the counterterm $\delta Z$ being defined in the modified
minimal subtraction scheme. We insert Eq.~(\ref{dz}) into the
expressions of the LO and NLO quark diagrams, and of the NLO
effective diagrams. The LO hard kernel $H^{(0)}$ multiplied by
$\delta Z$ then regularizes the ultraviolet pole in Eq.~(\ref{pgt}).
The ultraviolet poles in Eqs.~(\ref{bpt}) and (\ref{ppt}) are regularized by the
counterterm of the quark field and by an additive counterterm in the
modified minimal subtraction scheme.

The NLO hard kernel for Fig.~\ref{leading}(a) is given by
\begin{eqnarray}
H^{(1)} & = & { \alpha_s (\mu_f) C_F \over 4 \pi} \bigg [  {21
\over 4} \ln {\mu^2 \over m_B^2}  -\left ( \ln { m_B^2
\over\zeta_1^2 }  +{13 \over 2}\right) \ln {\mu_f^2 \over m_B^2}
+{9 \over 16}  \left (\ln^2 x_1 + 2 \ln x_1 \ln x_2 -\ln x_2^2
\right)  \nonumber \\
&& +  \left (2 \ln { m_B^2 \over\zeta_1^2 } +{7 \over 8} \ln \eta
- {1 \over 4} \right) \ln x_1 +  \left (2 \ln { m_B^2
\over\zeta_2^2 } +{7 \over 8} \ln \eta - {5 \over 2} \right) \ln
x_2 + 2  \ln { m_B^2 \over\zeta_2^2 } + \left ({15 \over 4} -{7
\over 16} \ln \eta \right ) \ln \eta  \nonumber \\
&& -{1 \over 2} \ln { m_B^2 \over\zeta_1^2 }   \left ( 3 \ln {
m_B^2 \over\zeta_1^2 } + 2 \right ) +{85 \over 48} \pi^2 +{219
\over 16} \bigg ] H^{(0)}, \label{pht}
\end{eqnarray}
in which all the infrared regulators $m_g$, $\delta_1$, and
$\delta_2$ have disappeared. A choice of the
scales $\zeta_1$ and $\zeta_2$ corresponds to a factorization
scheme, which should be fixed for consistency. Following the scheme
$\zeta^2=Q^2$ adopted in the NLO analysis of the pion transition
and electromagnetic form factors
\cite{NL07,LSW11}, we set $\zeta_2$ to $m_B^2$. The important logarithms
$\ln(m_B^2 /\zeta_1^2)$, arising from the $B$ meson wave
function, enter the hard kernel after the infrared subtraction. Instead of
performing resummation of these logarithms, we choose a
sufficiently large $\zeta_1$, say, $\zeta_1 / m_B =25$ in the
numerical analysis, which has been assumed for achieving the
simplified result in Eq.~(\ref{bwe}). In this scheme the
$\ln^2(m_B/\zeta_1)$ term happens to cancel the large constant
term in the hard kernel, and reduces the NLO correction.

Moreover, the double logarithm $\ln^2 x_2$ has been absorbed
into the jet function $J(x_2)$ \cite{UL} defined in
the kinematic region where the virtual $b$
quark in Fig.~\ref{leading}(a) becomes almost on-shell, namely, with
$x_2\to 0$. The organization of this important logarithm
to all orders leads to the threshold resummation factor,
which further suppresses the end-point singularity from small
$x_2$ in the $B\to\pi$ form factors \cite{TLS}.
Therefore, we have to subtract the NLO jet function
\cite{NL07}
\begin{eqnarray}
J^{(1)}H^{(0)}=-\frac{\alpha_s}{4\pi}C_F\left(
\ln^2 x_2+\ln x_2+\frac{\pi^2}{3}\right)H^{(0)},\label{pja}
\end{eqnarray}
from Eq.~(\ref{pht}), which finally turns into
\begin{eqnarray}
H^{(1)}&\to & H^{(1)}- J^{(1)}H^{(0)}\nonumber\\
& = & { \alpha_s (\mu_f) C_F \over 4 \pi} \bigg [  {21
\over 4} \ln {\mu^2 \over m_B^2}  -\left ( \ln { m_B^2
\over\zeta_1^2 }  +{13 \over 2}\right) \ln {\mu_f^2 \over m_B^2}
+{7 \over 16} \ln^2( x_1 x_2)+{1 \over 8}\ln^2 x_1  +{1 \over 4}\ln x_1\ln x_2 \nonumber \\
&& +  \left (2 \ln { m_B^2 \over\zeta_1^2 } +{7 \over 8} \ln \eta
- {1 \over 4} \right) \ln x_1 +  \left ({7 \over 8} \ln \eta - {3 \over 2} \right) \ln
x_2  + \left ({15 \over 4} -{7
\over 16} \ln \eta \right ) \ln \eta  \nonumber \\
&& -{1 \over 2} \ln { m_B^2 \over\zeta_1^2 }   \left ( 3 \ln {
m_B^2 \over\zeta_1^2 } + 2 \right ) +{101 \over 48} \pi^2 +{219
\over 16} \bigg ] H^{(0)}. \label{pht2}
\end{eqnarray}
Because the double logarithm $\ln^2 x_1$ was not resummed in
\cite{TLS}, it is left in the above NLO hard kernel $H^{(1)}$. Another
double logarithm $\ln^2(x_1 x_2)$ actually arises from the approximation
$\ln^2\delta_{12}\approx \ln^2(x_1 x_2)$. This approximation makes sense:
a logarithm does not develop an end-point singularity, so the
$k_T^2$ term is negligible in $\ln\delta_{12}$.
Equation~(\ref{pht2}), proportional to ${P_2}^\mu$, generates the NLO
corrections at leading twist to the $B\to\pi$ transition form factors
$f^+(q^2)$ and $f^0(q^2)$ in Eq.~(\ref{form}).

\section{NUMERICAL ANALYSIS}

In this section we evaluate the $B\to\pi$ transition form factors
numerically in the $k_T$ factorization up to NLO, adopting the
following non-asymptotic pion distribution amplitudes
\cite{Duplancic:2008ix,Khodjamirian:2011ub},
\begin{eqnarray}
 \phi_{\pi}^A(x) &=& \frac{6f_{\pi}}{2\sqrt{2N_c}} x(1-x)
 \left[ 1 +a_2 C_2^{3/2}(u)+a_4 C_4^{3/2}(u) \right],
 \nonumber\\
 \phi_{\pi}^P(x) &=& \frac{f_{\pi}}{2\sqrt{2N_c}}\left[1 +0.59 C_2^{1/2}(u) +0.09 C_4^{1/2}(u) \right], \nonumber\\
 \phi_{\pi}^{\sigma}(x) &=& \frac{6 f_{\pi}}{2\sqrt{2N_c}} x (1-x) \left[1+0.11 C_2^{3/2}(u)\right],\label{pion: non
 asy}
\end{eqnarray}
with the pion decay constant $f_\pi=130$ MeV, the Gegenbauer moments
$a_2=0.16$ and $a_4=0.04$, and the Gegenbauer polynomials
\begin{eqnarray}
 &C^{1/2}_{1}(u)=u ,&C^{3/2}_{1}(u)=3u,  \nonumber\\
 &C_2^{1/2}(u)=\frac{1}{2} (3u^2-1),& C_2^{3/2} (u)=\frac{3}{2}
(5u^2-1),\nonumber \\
&C_3^{1/2} (u) = \frac{1}{2} u (5u^2 -3), & C_4^{3/2}
(u)=\frac{15}{8} (21 u^4 -14 u^2 +1),
\end{eqnarray}
and the variable $u=1-2x$. The $B$ meson distribution amplitudes
inspired from  a QCD sum rule analysis in the HQET \cite{Grozin:1996pq}
\begin{eqnarray}
 \phi_{B}^{(+)}(x, b) &=& \frac{f_{B}}{2\sqrt{2N_c}}  \, \, x \left ({m_B \over \omega_0 } \right)^2 {\rm Exp}
 \left [- {x m_B \over \omega_0 } -{1 \over 2} (\omega_0 b)^2
 \right], \nonumber \\
 \phi_{B}^{(-)}(x, b) &=&  \frac{f_{B}}{2\sqrt{2N_c}}  \, \left ({m_B \over \omega_0 } \right) {\rm Exp}
 \left [- {x m_B \over \omega_0 } -{1 \over 2} (\omega_0 b)^2
 \right],
 \label{B meson DA}
\end{eqnarray}
are employed, where the $B$ meson decay constant is set to a
constant $f_B=214$ MeV for convenience (namely, neglecting its
evolution), and the shape parameter is chosen as $\omega_0=0.35$
GeV.

\begin{figure}[t]
\begin{center}
\includegraphics[height=7.5 cm]{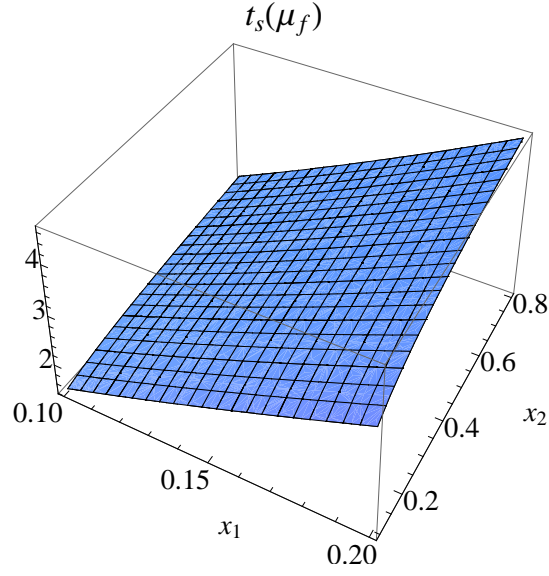}
\caption{Renormalization scale $t_s (\mu_{\rm f})$, defined in Eq.
(\ref{ts function}), as a function of momentum fractions $x_1$ and
$x_2$ for a typical factorization scale $\mu_f =1.5$ GeV and
the ratio ${\zeta_1 / m_B }=25$. }
\label{renormalization scale}
\end{center}
\end{figure}

\begin{figure}[t]
\begin{center}
\includegraphics[height=6.5 cm]{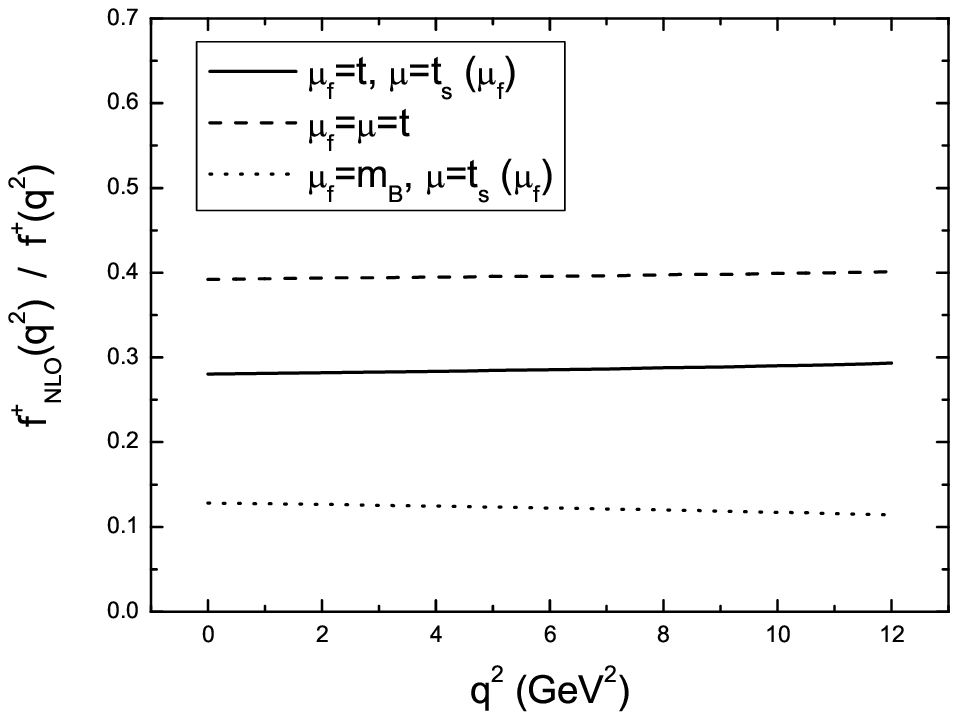} \hspace{0.2 cm}
\includegraphics[height=6.5 cm]{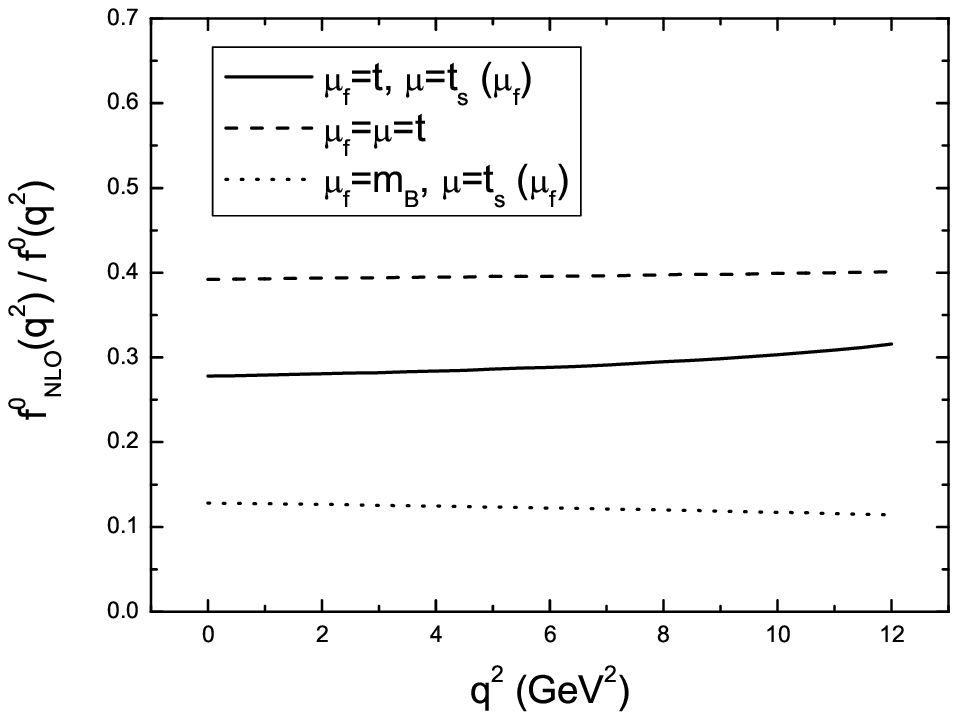}
\caption{Ratios of the NLO corrections over the total contributions
to the $B \to \pi$ form factors for three different choices of
the renormalization and factorization scales. } \label{ratio}
\end{center}
\end{figure}

The first issue concerns the choice of the renormalization scale
$\mu$ and the factorization scale $\mu_{\rm f}$ in order to
minimize the NLO corrections to the form factors. For the first
choice, $\mu_{\rm f}$ is set to the hard scales specified in the
PQCD approach to exclusive processes \cite{LS,KLS,b-baryon}
\begin{eqnarray}
t^{a}=\max(\sqrt{x_2 \eta } \, m_B ,1/b_1,1/b_2), \qquad
t^{b}=\max(\sqrt{x_1 \eta } \, m_B ,1/b_1,1/b_2),\label{tab}
\end{eqnarray}
corresponding to the largest energy scales in Figs.~\ref{leading}(a)
and \ref{leading}(b),  respectively. Then we utilize the freedom of
choosing $\mu$ to diminish all the single-logarithmic and constant
terms in the NLO hard kernel, which is found to be
\begin{eqnarray}
t_s (\mu_{\rm f})  = \left \{{\rm Exp} \left[ c_1 + \left(\ln
{m_B^2 \over \zeta_1^2}  +{5 \over 4} \right)  \ln{\mu_{\rm f}^2
\over m_B^2 } \right ]  \, x_1^{c_2 } \, x_2^{c_3} \right
\}^{2/21} \, \mu_{\rm f}, \label{ts function}
\end{eqnarray}
with the coefficients
\begin{eqnarray}
c_1 &=& - \left ({15 \over 4} -{7 \over 16} \ln \eta \right ) \ln
\eta + {1 \over 2} \ln { m_B^2 \over\zeta_1^2 }   \left ( 3 \ln {
m_B^2 \over\zeta_1^2 } + 2 \right ) - {101 \over 48} \pi^2 - {219
\over 16} \,,  \nonumber \\
c_2 &=& - \left ( 2 \ln { m_B^2 \over\zeta_1^2 }  + {7 \over 8}
\ln \eta - {1 \over 4} \right )  \,, \nonumber \\
c_3 &=& -{7 \over 8} \ln \eta + {3 \over 2}. \nonumber
\end{eqnarray}
To have an idea of the magnitude of the renormalization scale
$\mu=t_s (\mu_{\rm f})$, we display its behavior in the dominant
region with the small momentum fractions $x_1$ and $x_2$ in
Fig.~\ref{renormalization scale}, where the factorization scale
$\mu_f$ is fixed at its typical value 1.5 GeV.
The ratio of the NLO contributions over the total ones as a
function of the transfer momentum squared $q^2$ is summarized in
Fig.~\ref{ratio}, which is approximately 30\% for both the form
factors $f^+(q^2)$ and $f^0(q^2)$. In the second choice, we set
both scales to $\mu=\mu_{\rm f}=t^a$ ($t^b$) in the factorization
formula associated with Fig.~\ref{leading}(a) (\ref{leading}(b)).
It turns out that this simple scenario yields larger NLO
corrections around 40\% as shown in Fig.~\ref{ratio}. The third
choice corresponds to $\mu_{\rm f}=m_B$ and $\mu=t_s(\mu_{\rm
f})$, for which the NLO corrections are approximately 15\%.
However, the inverse relation $\mu_f > \mu$ in this case seems not
to be natural. Hereafter, we shall adopt the first choice of the
renormalization and factorization scales as the default one.

\begin{figure}[t]
\begin{center}
\includegraphics[height=6.5 cm]{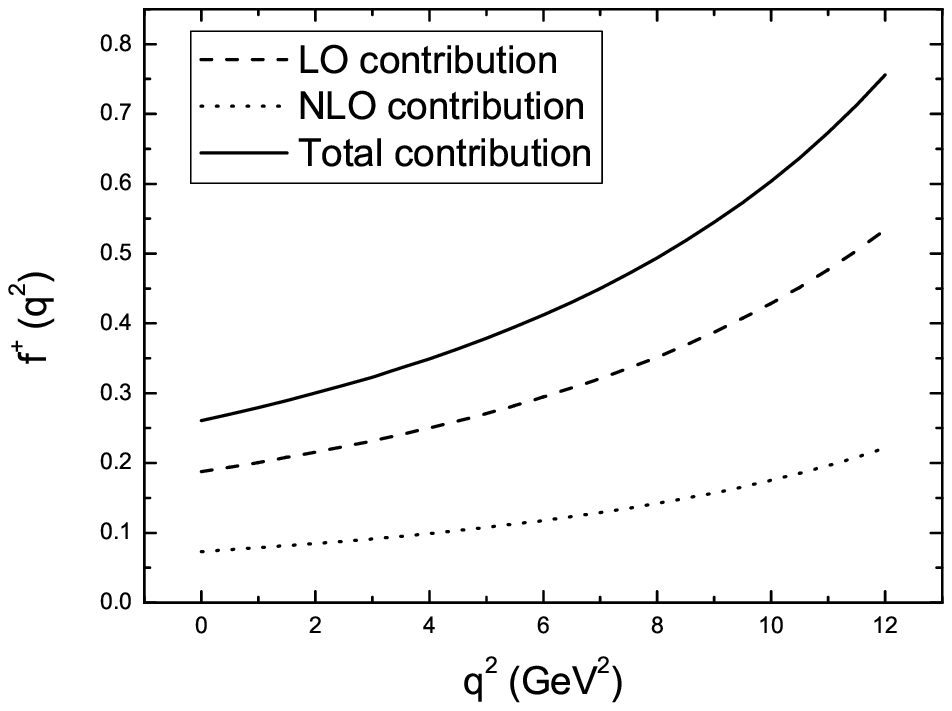} \hspace{0.2 cm}
\includegraphics[height=6.5 cm]{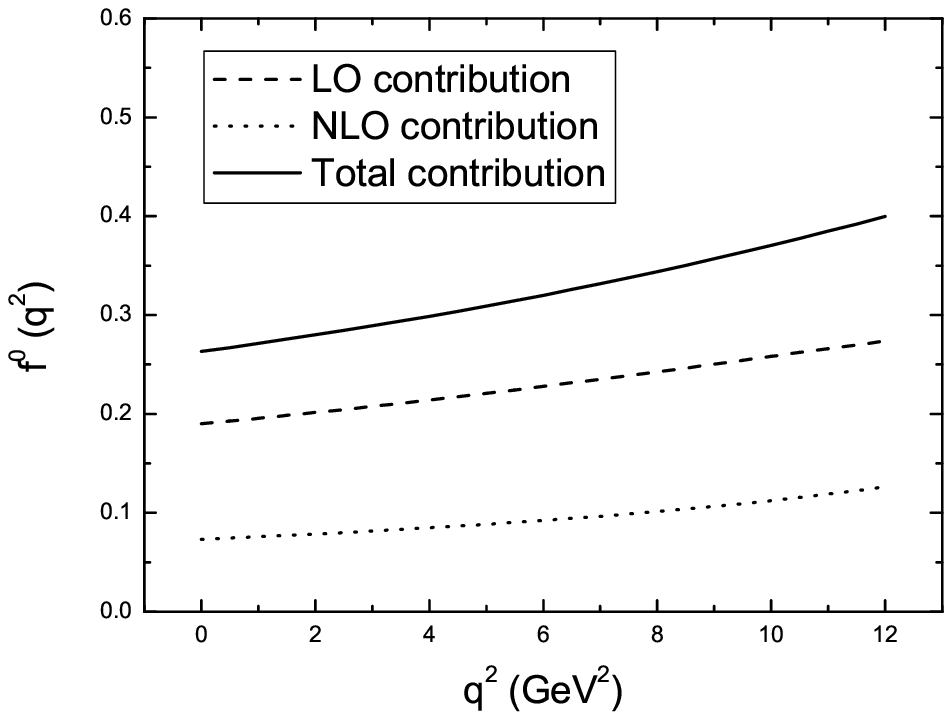}
\caption{LO and NLO contributions to the $B \to \pi$ form factors
with the non-asymptotic pion distribution amplitudes in
Eq.~(\ref{pion: non asy}) and the first scenario for the scale choice,
$\mu_{\rm f}=t$ and $\mu=t_s(\mu_{\rm f})$. } \label{B to pi form factor:
nonasymptotic Pion DA}
\end{center}
\end{figure}

The $B \to \pi$ transition form factors in the $k_T$ factorization up to NLO are
presented in Fig.~\ref{B to pi form factor: nonasymptotic Pion DA}.
It is observed that the LO and NLO contributions exhibit the similar
power-law behavior, as they should. It is not a surprise that the form
factors at the maximal recoil of the pion, $f^+(0)=f^0(0)$, are close to their LO
value \cite{TLS}, even after including the NLO contributions. The reason
is that the meson wave functions have been adjusted accordingly to
maintain this value, which is regarded as an input. That is, when
choosing hadron wave functions in the PQCD approach, one must pay
attention to the order in the coupling constant, at which the hadron
wave functions are determined. Though the form-factor values, treated as
inputs, are not changed at higher orders, the different hadron wave
functions extracted at different orders do affect
other topologies of nonleptonic two-body $B$ meson decay amplitudes.
It is worthwhile to investigate the corrections to nonleptonic two-body
$B$ meson decays from this NLO source in future works.

\begin{figure}[t]
\begin{center}
\includegraphics[height=6.5 cm]{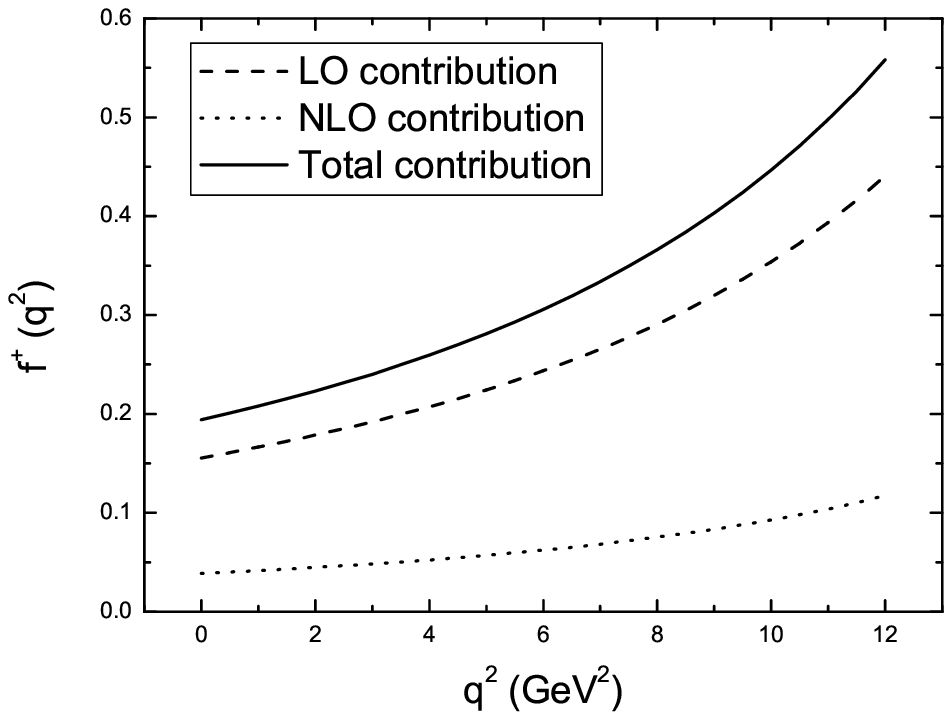} \hspace{0.2 cm}
\includegraphics[height=6.5 cm]{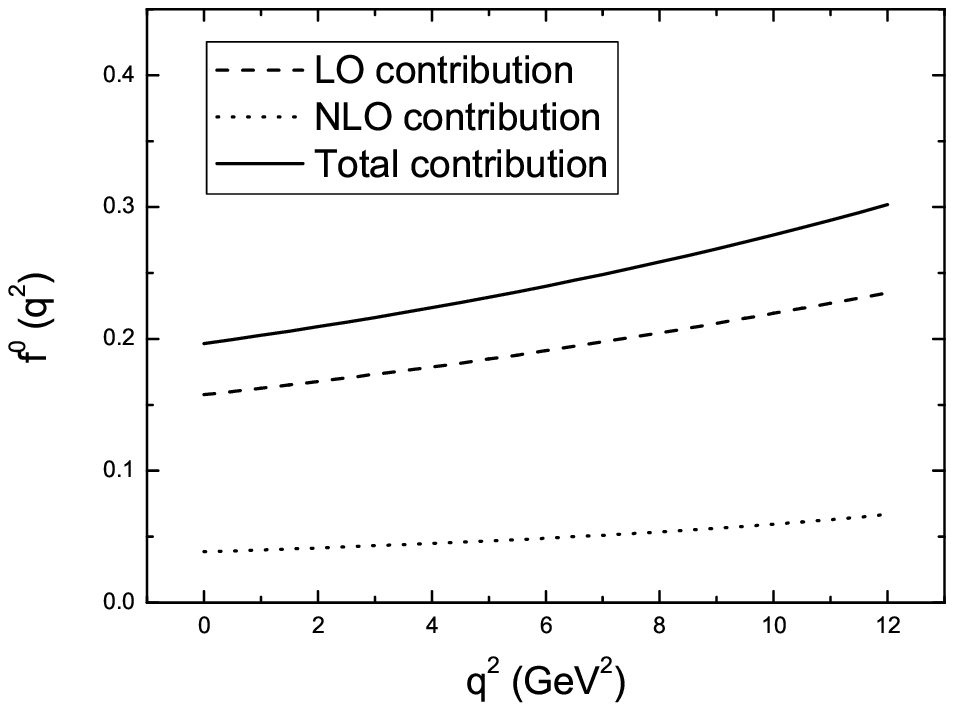}
\caption{LO and NLO contributions to the $B \to \pi$ form factors
with the asymptotic pion distribution amplitudes and the first
scenario for the scale choice,
$\mu_{\rm f}=t$ and $\mu=t_s(\mu_{\rm f})$. }
\label{B to pi form factor: asymptotic Pion DA}
\end{center}
\end{figure}

\begin{figure}[t]
\begin{center}
\includegraphics[height=6.0 cm]{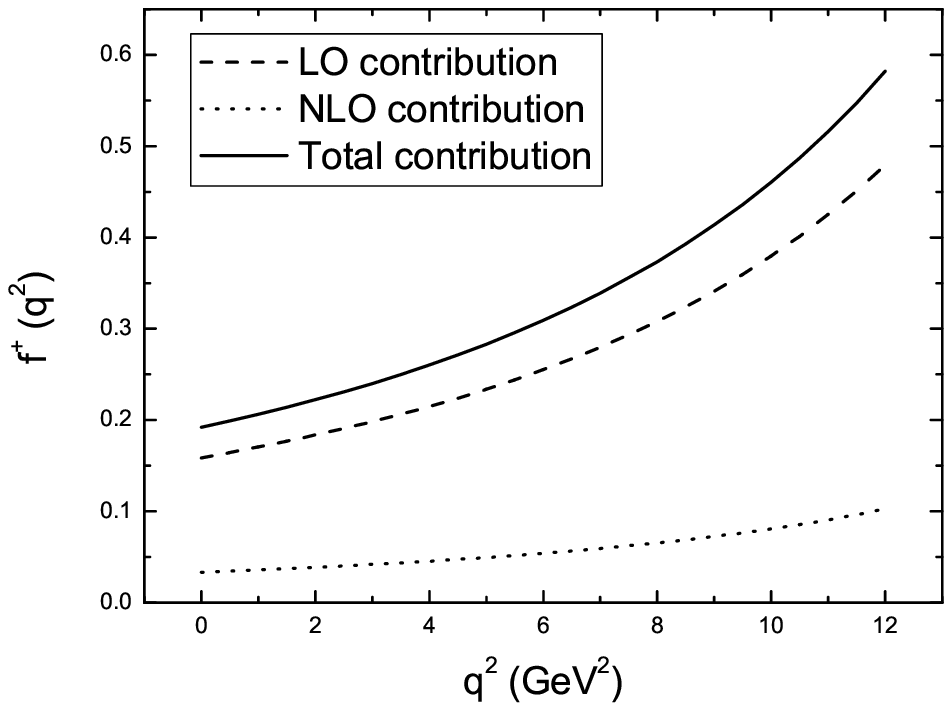} \hspace{0.2 cm}
\includegraphics[height=6.0 cm]{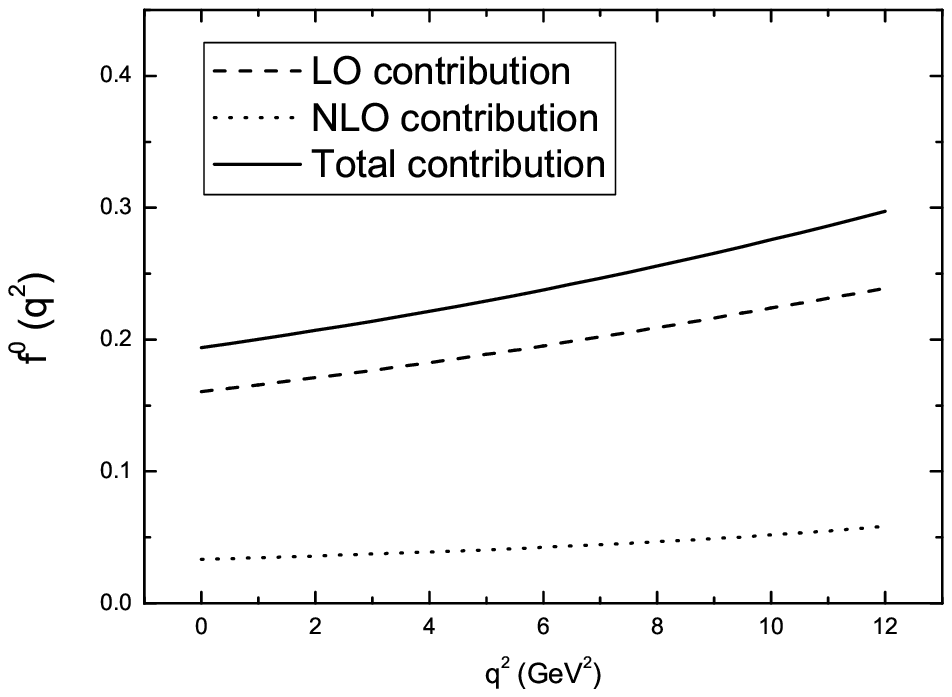}
\caption{LO and NLO contributions to the $B \to \pi$ form factors
with the  $B$ meson distribution amplitudes in Eq.~(\ref{B meson
DA:PQCD}) and the first scenario for the scale choice: $\mu_f=t$ and
$\mu=t_s(\mu_f)$. } \label{B to pi form factor: kT B meson DA}
\end{center}
\end{figure}

To test the impact of higher conformal-spin partial waves in the
pion distribution amplitudes, we plot the $q^2$ dependence of the
form factors in Fig.~\ref{B to pi form factor: asymptotic Pion DA}
with the asymptotic pion distribution amplitudes. Numerically,
both the form factors are reduced by about 25\% for $q^2 \leq 12
\, {\rm GeV}^2$ without the non-asymptotic Gegenbauer terms in
Eq.~(\ref{pion: non asy}). We also investigate the effects from
different models of the $B$ meson distribution amplitudes. A model
widely adopted in the PQCD analysis is given by
\begin{eqnarray}
\phi_{B}^{(+)}(x, b) = \phi_{B}^{(-)}(x, b) =
\frac{f_{B}}{2\sqrt{2N_c}} \,  N_B \,  x^2 (1-x)^2 \, {\rm Exp}
\left [- {x^2 m_B \over 2 \omega_0^2 } -{1 \over 2} (\omega_0 b)^2
\right], \label{B meson DA:PQCD}
\end{eqnarray}
with the normalization constant $N_B$ defined via $\int
dx\phi_{B}^{(+)}(x, 0)=f_B/(2\sqrt{2N_c})$. Note that the two
leading $B$ meson distribution amplitudes have been assumed to be
equal for the purpose of numerical estimate, which do not obey the
equations of motion \cite{KKQT}. Besides, it exhibits an asymptotic
behavior at $x \to 0$ different from that derived in
\cite{Lange:2003ff}. The corresponding $q^2$ dependence in
Fig.~\ref{B to pi form factor: kT B meson DA} indicates that the
form factors with the model in Eq.~(\ref{B meson DA:PQCD}) are
approximately 25\% smaller than those with the model in Eq.~(\ref{B
meson DA}). It is interesting to notice in Fig.~\ref{B to pi form
factor: kT B meson DA} that the NLO corrections are relatively
small, less than 20\% of the total contributions. The reason is
attributed to the fact that the end-point region of $x_1$ is
strongly suppressed by this model and the double logarithm $\ln^2
x_1$ in the NLO hard kernel does not play an essential role.

\begin{figure}[t]
\begin{center}
\includegraphics[height=6.5 cm]{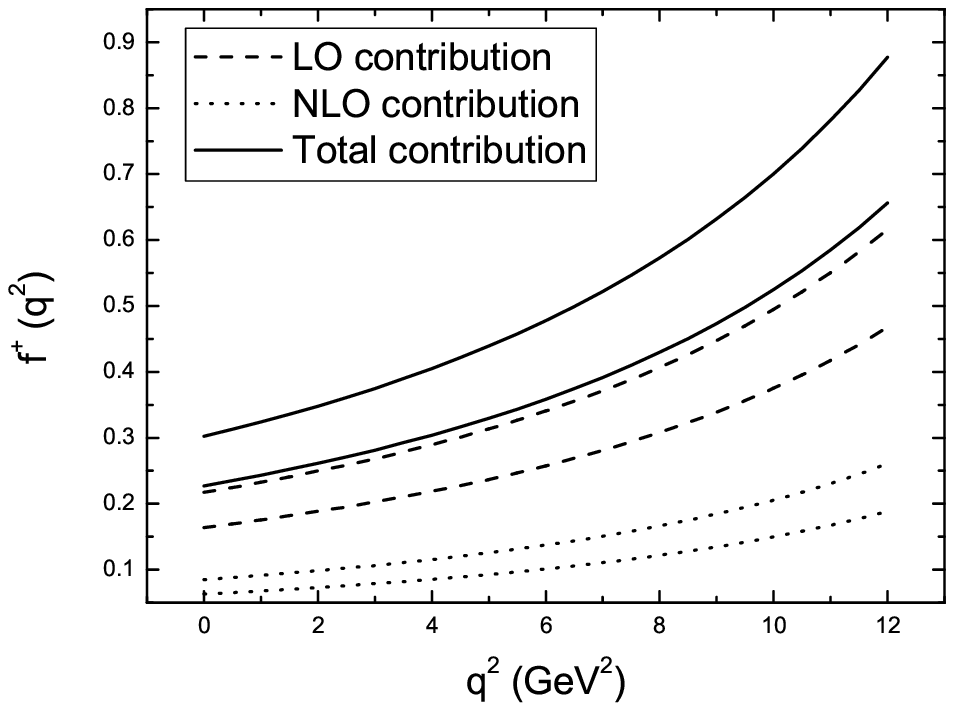} \hspace{0.2 cm}
\includegraphics[height=6.5 cm]{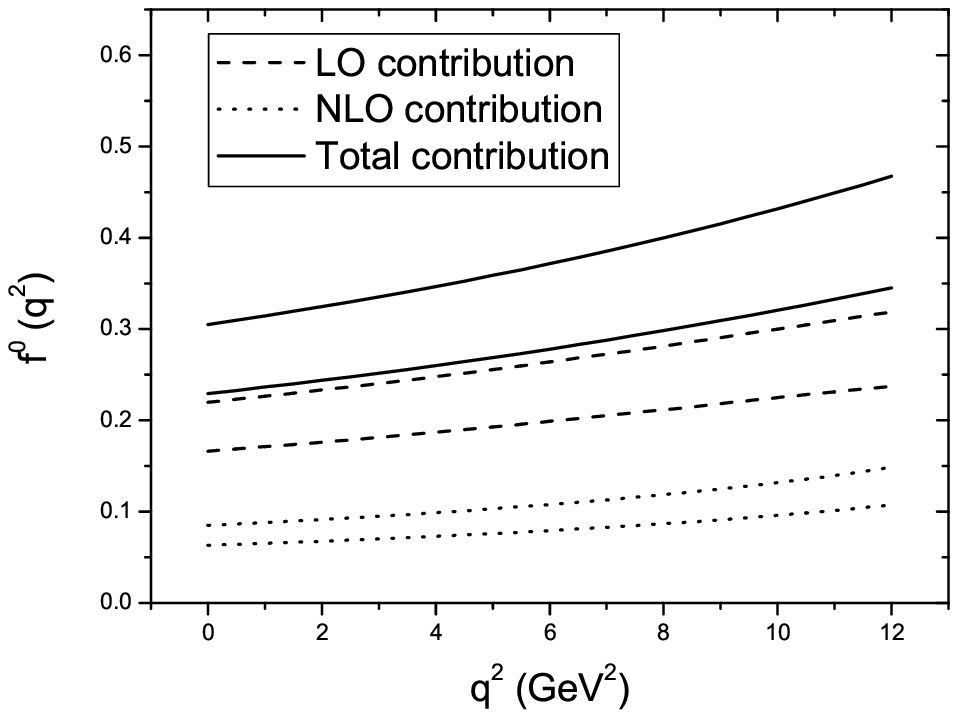}
\caption{LO and NLO contributions to the $B \to \pi$ form factors
with the  $B$ meson distribution amplitudes in Eq.~(\ref{B meson
DA}), however,  varying  the shape parameter $\omega_0$ from
$0.30$ GeV to $0.40$ GeV and the first scenario for the scale
choice: $\mu_f=t$ and $\mu=t_s(\mu_f)$. } \label{B to pi form
factor: varying omega0}
\end{center}
\end{figure}

\begin{figure}[t]
\begin{center}
\includegraphics[height=6.5 cm]{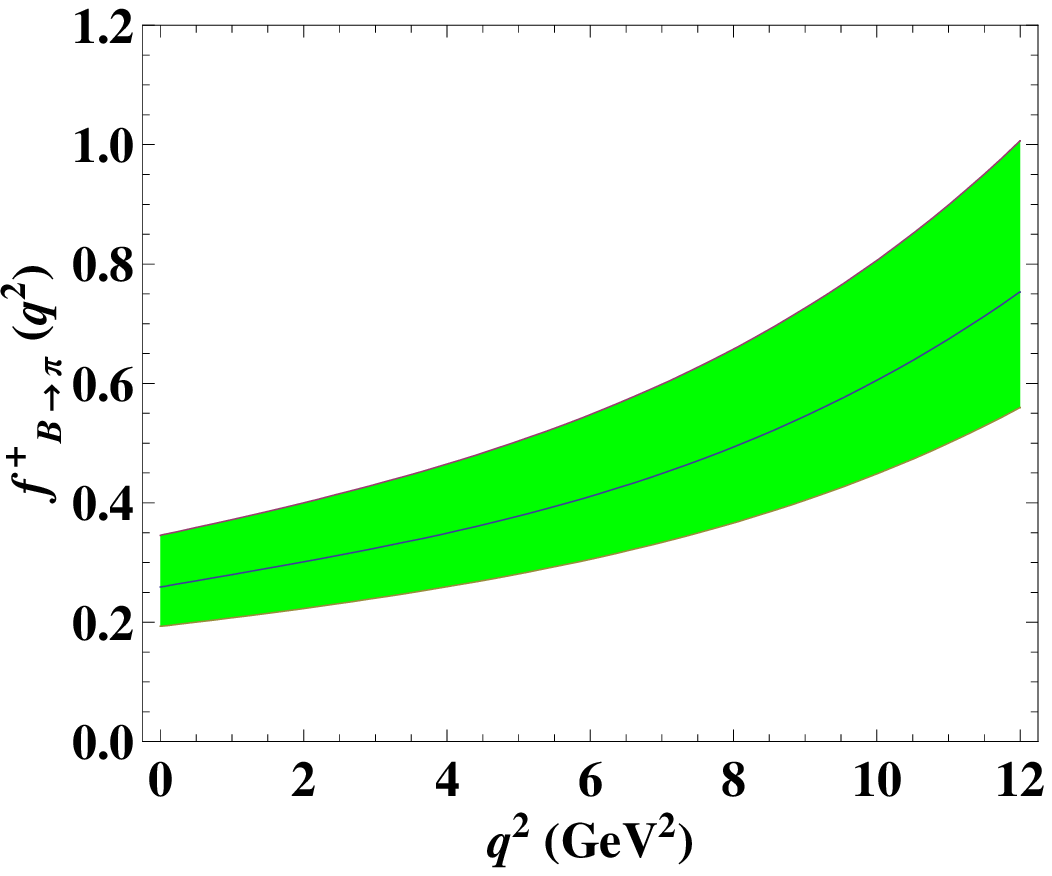} \hspace{0.2 cm}
\includegraphics[height=6.5 cm]{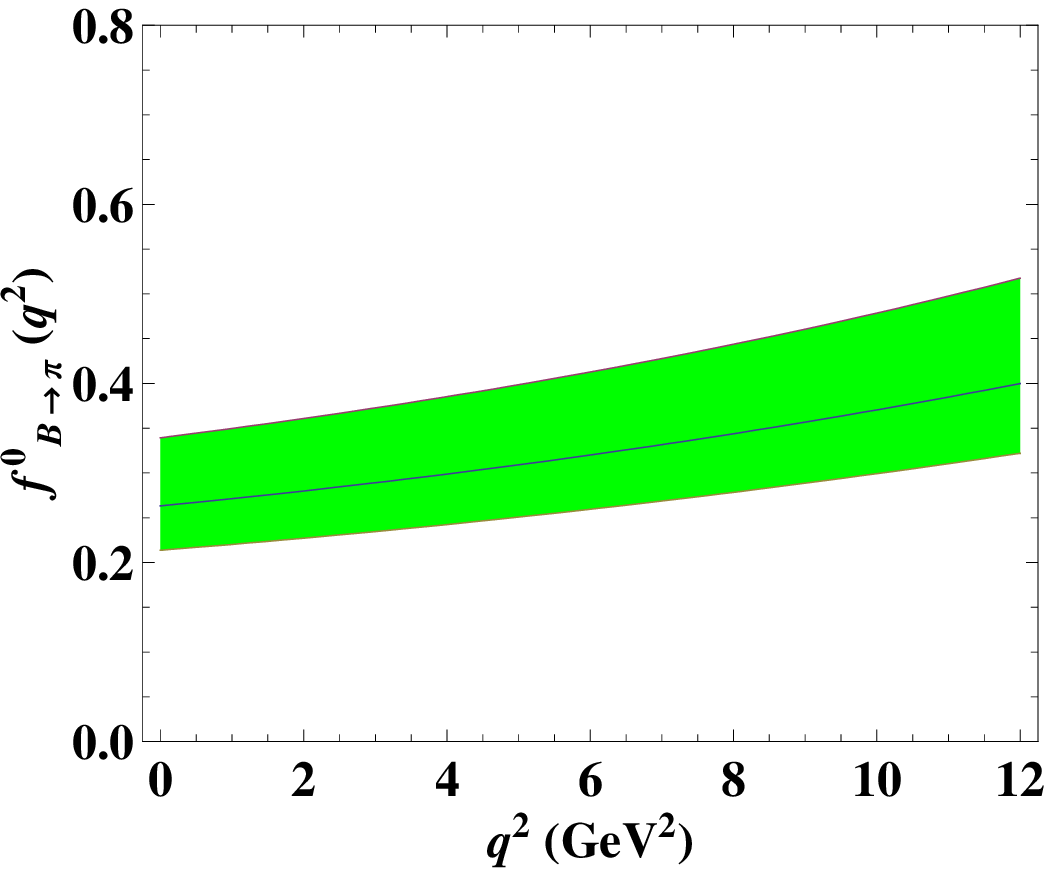}
\caption{Theoretical uncertainties of  the $B \to \pi$ form
factors with the first scenario for the scale choice: $\mu_f=t$
and $\mu=t_s(\mu_f)$. } \label{B to pi form factor: error}
\end{center}
\end{figure}

The extraction of the Cabibbo-Kobayashi-Maskawa matrix element
$|V_{ub}|$ \cite{KM} from the semileptonic decay
$B\to\pi\ell\bar\nu$ is of intensive phenomenological interest
recently (see \cite{Khodjamirian:2011ub} and references therein).
Here we comment on the consistency of the $B\to\pi$ form factors
predicted in the NLO $k_T$ factorization with those in the
literature, in view of the extraction of $|V_{ub}|$. For this
purpose, we also estimate the theoretical uncertainties of the
form factors $f^{+} (q^2)$ and $f^{0} (q^2)$ from the variations
of the Gegenbauer moments $a_2$ and $a_4$ in the twist-2 pion
distribution amplitudes,  from the variations of the chiral scale
$m_0\equiv m_{\pi}^2/(m_u+m_d)$, $m_u$ ($m_d$) being the $u$ ($d$)
quark mass, involved in the two-parton twist-3 pion distribution
amplitudes \cite{PB2}, and from the variations of the shape
parameter $\omega_0$ in the $B$ meson distribution amplitudes in 
Eq.~(\ref{B meson DA}).
Unfortunately, the extraction of  $\omega_0$  still suffers large
uncertainty from QCD sum-rule calculations. We simply
take $\omega_0=0.35 \pm 0.05 \, {\rm GeV}$ to illustrate the effect 
on the form factors from the variation of $\omega_0$. It
is seen from Fig.~\ref{B to pi form factor: varying omega0}
that both the form factors $f^{+} (q^2)$ and $f^{0} (q^2)$,
including LO and NLO contributions, 
increase (decrease) by 15 \% with the decrease (increase) of 
$\omega_0$. Combining the uncertainties due to $a_2 ( {1 \rm
GeV})=0.16^{+0.09}_{-0.07}$, $a_4 ( {1 \rm GeV})
=0.04^{+0.12}_{-0.08}$,  $m_0 ( {1 \rm GeV}) =1.74^{+0.67}_{-0.38}
\, {\rm GeV}$, and $\omega_0=0.35 \pm 0.05 \, {\rm GeV}$, we
predict the  form factors $f^{+} (q^2)$ and $f^{0} (q^2)$ as
displayed in Fig.~\ref{B to pi form factor: error}. Fitting to the
BaBar data on the integrated $B\to\pi\ell\bar\nu$ branching ratio
within the region $0 \leq q^2 \leq 8$ GeV$^2$ \cite{:2010uj},
where the leading-twist $k_T$ factorization is expected to work
well, we obtain
\begin{eqnarray}
|V_{ub}|= 2.90^{+0.77}_{-0.80} \big|_{th.} {}^{+0.13}_{-0.14}
\big|_{exp.}. \label{Vub result}
\end{eqnarray}
The above value is in good agreement with that in \cite{:2010uj},
which employed the data on $q^2$ bins in the whole kinematic
region and the lattice QCD results of the $B\to\pi$ form factors
from the FNAL/MILC Collaboration \cite{Cundy:2009yy}.
Equation~(\ref{Vub result}), however, differs from $|V_{ub}|=
3.59^{+0.38}_{-0.33} \big|_{th.}\pm 0.11 \big|_{exp.}$ extracted
in \cite{Khodjamirian:2011ub}, where the $B\to\pi$ form factors
were computed in the light-cone sum rule (LCSR). The distinction
can be traced back to the different $q^2$ dependence of the form
factor $f^{+} (q^2)$ predicted in the $k_T$ factorization and in
LCSR, albeit with the similar $f^{+} (0)$ value in both
approaches. More dedicated efforts on the study of the shape of $B
\to \pi$ form factors in QCD is in demand in order to resolve 
the potential difference in the extraction of $|V_{ub}|$.

\section{CONCLUSION}

In this paper we have calculated the NLO corrections to the
$B\to\pi$ transition form factors at leading twist in the $k_T$
factorization theorem. Both the collinear and soft divergences in
the NLO quark diagrams and in the NLO effective diagrams for meson
wave functions are regularized by the off-shellness $k_T^2$ of
light partons. The $b$ quark remains on-shell, such that it can be
approximated by the standard effective heavy quark in the $k_T$
factorization. The key is that the soft gluons radiated by the $b$
quark and attaching to other particle lines can be regularized by
the virtuality of other particle lines. The NLO pion wave function
is the same as constructed in the pion transition and
electromagnetic form factors, confirming its universality.
Compared to the pion wave function, the NLO $B$ meson wave
function contains the additional double logarithm
$\ln^2(\zeta_1^2/m_B^2)$. Because of the assumed hierarchy
$\zeta_1^2\gg m_B^2$, the appearance of this double logarithm
demands the implementation of the resummation technique, which is
expected to minimize the scheme dependence from different choices
of $\zeta_1$. This subject, together with the asymptotic behavior
of the $B$ meson wave function in the $k_T$ factorization, will be
discussed in a forthcoming work.

The exact cancellation of the infrared divergences between the
quark diagrams and the effective diagrams verifies the validity of
the $k_T$ factorization for the $B$ meson semileptonic decays at
NLO level. Though the NLO hard kernel for the $B\to\pi$ transition form
factors contains a huge constant term, it is reduced by the
large double logarithm $\ln^2(\zeta_1^2/m_B^2)$ mentioned above.
This is the reason why the conventional choice of the
factorization scale in the PQCD approach, as the virtuality of internal
particles, can work to render the NLO corrections under control. By
tuning the renormalization scale to cancel the
single-logarithmic and constant terms, which is still lower than
the $B$ meson mass in the dominant kinematic region, the NLO
corrections are about 30\% of the form factors. The effect of
varying the meson wave functions has been also investigated:
the model for the $B$ meson wave function with a
stronger suppression at a small momentum fraction, and the
asymptotic model for the pion wave function lower the NLO
corrections down to 20\%.

\section*{Acknowledgement}

We are grateful to  Thomas Mannel for helpful discussions. The
work was supported in part by the National Science Council of
R.O.C. under Grant No. NSC-98-2112-M-001-015-MY3, by the National
Center for Theoretical Sciences of R.O.C., by National Science
Foundation of China under Grant No. 11005100, by the German
research foundation DFG under contract MA1187/10-1, and by the
German Ministry of Research (BMBF) under contract 05H09PSF.

\end{document}